\documentclass[12pt,preprint]{aastex}

\usepackage{natbib}
\usepackage{graphicx}
\newcommand{\hm}{H$_{2}$}

\shorttitle{CO depletion}
\shortauthors{Cazaux et al.}

\begin{document}

\title{CO depletion: a microscopic perspective}

\author{S. Cazaux\altaffilmark{1,2,3}, R. Mart{\'{\i}}n-Dom{\'e}nech\altaffilmark{4},  Y. J. Chen\altaffilmark{5}, G. M. Mu{\~n}oz Caro\altaffilmark{4},  C. Gonz{\'a}lez D{\'i}az \altaffilmark{4}} 
\altaffiltext{1}{Faculty of Aerospace Engineering, Delft University of Technology, Delft, The Netherlands}
\altaffiltext{2}{University of Leiden, PO Box 9513, NL, 2300 RA, Leiden, The Netherlands}
\altaffiltext{3}{Kapteyn Astronomical Institute, University of Groningen, PO Box 800, 9700 AV, Groningen, The Netherlands}
\altaffiltext{4}{Centro de Astrobiolog\'ia (INTA-CSIC), Ctra. de Ajalvir, km 4, Torrej\'on de Ardoz, 28850 Madrid, Spain}
\altaffiltext{5}{Department of Physics, National Central University, Jhongli City, 32054, Taoyuan County, Taiwan}


\begin{abstract}

 In regions where stars form, variations in density and temperature can cause gas to freeze-out onto dust grains forming ice mantles, which influences the chemical composition of a cloud. The aim of this paper is to understand in detail the depletion (and desorption) of CO on (from) interstellar dust grains. Experimental simulations were performed under two different (astrophysically relevant) conditions. In parallel, Kinetic Monte Carlo simulations were used to mimic the experimental conditions. In our experiments, CO molecules accrete onto water ice at temperatures below 27 K, with a deposition rate that does not depend on the substrate temperature. During the warm-up phase, the desorption processes do exhibit subtle differences indicating the presence of weakly bound CO molecules, therefore highlighting a low diffusion efficiency. IR measurements following the ice thickness during the TPD confirm that diffusion occurs at temperatures close to the desorption. Applied to astrophysical conditions, in a pre-stellar core, the binding energies of CO molecules, ranging between 300 K and 850 K, depend on the conditions at which CO has been deposited. Because of this wide range of binding energies, the depletion of CO as a function of AV is much less important than initially thought. The weakly bound molecules, easily released into the gas phase through evaporation, change the balance between accretion and desorption, which result in a larger abundance of CO at high extinctions. In addition, weakly bound CO molecules are also be more mobile, and this could increase the reactivity within interstellar ices.
 \end{abstract}

\keywords{dust, extinction - ISM: abundances - ISM: molecules - stars: formation }

\maketitle

\section{Introduction}
In the last decades, observing facilities have significantly increased in sensitivity allowing to study in detail the chemical 
composition of many places of our Universe. Molecules and atoms are powerful indicators of the gas characteristics of a medium 
and are used to derive detailed properties of astrophysical objects. 
\rm{In particular,} observations of star forming environments have rapidly been confronted with the impossibility to explain 
the abundances of some species with gas-phase reactions only. In regions where stars \rm{form}, about 1\% of the mass is 
constituted by small dust particles ranging in size from few tens of \AA\ to a few micrometer. However small and inconspicuous 
these dust grains seem, they interact with the gas phase and can dramatically alter its composition. 
In the first phases of star formation, the \rm{molecular} clouds present some overly dense regions, called pre-stellar cores, 
which are the precursors of the stars. To reproduce the observations \rm{of dense clouds}, about 90\% of CO molecules should 
leave the gas phase, on average along the line of sight, and over 99\% of them must deplete in the core nucleus 
(\citealt{caselli1999}). This is due to CO freeze-out onto dust particles, which then form thick icy mantles (e.g. 
\citealt{ossenkopf1994}; \citealt{pontoppidan2008}). 
When the protostar forms and heats its surroundings, a rich molecular 
chemistry is triggered driven by thermal desorption of \rm{the ice mantles} (\citealt{cazaux2003}). This chemically rich phase 
is called hot core/hot corino (for high/low mass stars) and characterised by an abundant organic inventory (water and organics 
such as H$_2$CO and CH$_3$OH (\citealt{schoier2002}), complex O- and N-bearing molecules such as formic acid and acetaldehyde (\citealt{cazaux2003})). 
In order to explain these observations, the understanding of CO interaction with dust surfaces is unavoidable. 

In this study we followed experimentally the formation of CO ices deposited at different \rm{conditions}. 
For this purpose, we \rm{performed} two experimental simulations focusing on two different accretion processes: 
1) accretion onto a water ice substrate at a decreasing temperature, from 80 K to 8 K, and 
2) accretion at a constant temperature of 14 K. 
After accretion, the temperature of the \rm{substrate} was increased \rm{in both cases}, and CO molecules evaporated and 
could be measured in the gas phase. This is called a temperature programmed desorption (TPD) experiment. 
The present study \rm{aims to} understand whether TPD experiments can be sensitive to the different accretion processes. 
\rm{In this work, we show that subtle deviation between TPDs are crucial to constrain the binding energies and the diffusion of CO molecules in the ice. Furthermore, combining quadrupole mass spectrometry, infrared spectroscopy and laser interference allows to follow simultaneously the solid and gaseous phases of the CO molecules during the TPD and further constrain the desorption and diffusion processes.}  \rm
The experimental results are supported by theoretical calculations taking into account the microphysics occurring in ices. Our results are then exported to astrophysical conditions. 

This paper is articulated as follow. In section 2 the two experiments performed in this study are described as well as their results. 
In section 3, the theoretical model and assumptions are described and used to reproduce the experimental results. 
In section 4 the model is extended to pre-stellar cores in order to reproduce the CO depletion observed in these objects. 

\section{Experiments}
Deposition \rm{followed by} warming up of CO ices has been studied experimentally in two different setups, \rm{under two different 
conditions} in order to address whether the deposition conditions influence the \rm{subsequent thermal} desorption of the CO ices. 
The two experimental setups are described in the following sections, as well as the experimental results.

\begin{figure*} 
    \centering
    \includegraphics[width=0.55\textwidth]{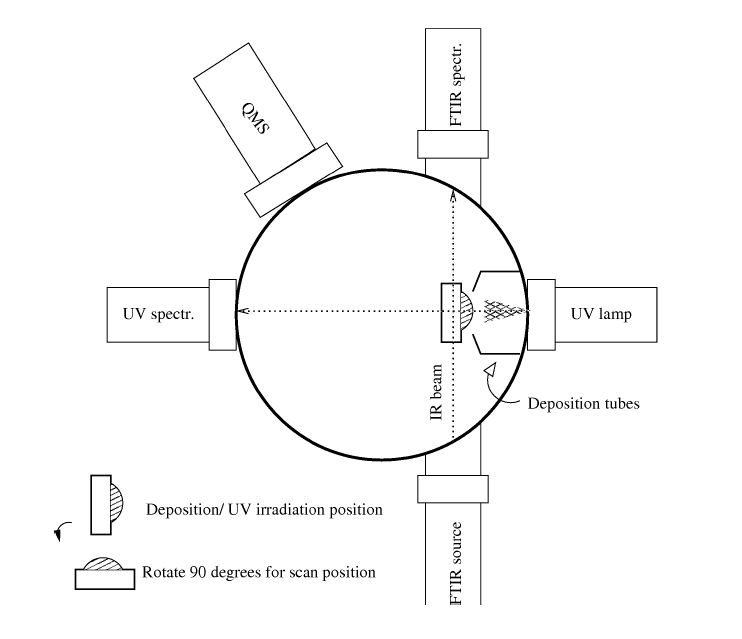}
    \caption{Schematic representation of the upper level of the main chamber of the ISAC experimental set-up, where gas deposition onto the cold substrate forms an ice layer that is UV irradiated. FTIR and QMS techniques allowed in situ monitoring of the solid and gas phases, from \cite{munozcaro2010}.}
    \label{setup2}
\end{figure*}

\subsection{
Experimental setup: the InterStellar Astrochemistry Chamber}
\rm{The experimental simulations corresponding to the accretion of CO molecules at a decreasing temperature onto a previously 
deposited water ice substrate were} 
performed using the InterStellar Astrochemistry Chamber (ISAC) at the Centro de Astrobiolog\'ia. 
\rm{The aim of these experiments was to determine the maximum temperature at which CO molecules were able to accrete onto 
the water ice and form an ice mantle.}
ISAC is an ultra-high vacuum (UHV) chamber with pressure typically in the range P = 2.5-4.0 $\times$ 10$^{-11}$ mbar, 
which corresponds to dense cloud interiors. A \rm{schematic representation} of this setup is shown in figure~\ref{setup2}. 
The chemical components used in the experiments were H$_2$O (vapor, triply distilled), and CO (gas), 
that were introduced into the chamber from an independent gas-line system through a capillary tube, condensing onto a 
KBr substrate and forming an ice analog. 
\rm{A closed-cycle helium cryostat and a tunable heater, combined with} a silicon diode temperature sensor and a LakeShore Model 331 
temperature controller were used \rm{to control the sample temperature}, reaching a sensitivity of about 0.1 K. 
The evolution of the solid sample was monitored by in situ Fourier transform infrared (FTIR) spectroscopy in transmittance (model Bruker 
Vertex 70, equipped with a deuterated triglycine sulfate detector, or DTGS), with a spectral resolution of 2 cm$^{-1}$ (\citealt{munozcaro2010}). 
Column densities of each species in the ice were calculated from the IR spectra using the formula
\begin{equation}
N=\frac{1}{A}\int_{\rm{band}} \tau_{\nu} \ d\nu
\end{equation}
where $N$ is the column density in molecules cm$^2$, $\tau_{\nu}$ the optical depth of the absorption band, and $A$ the band 
strength in cm molecule$^{-1}$, where we adopt a value of $A$(CO)=1.1 $\times$ 10$^{-17}$ cm molecule$^{-1}$ (\citealt{jiang1975})
and \rm{A(H$_2$O)=2.0 $\times$ 10$^{-16}$ cm molecule$^{-1}$ (\citealt{hagen1981})}.
\rm{A total of ~35 ML (1 ML = 10$^{15}$ molecules cm$^{-2}$) of amorphous solid water (ASW) were first deposited onto the KBr substrate at 80 K with an accretion rate of ~6 ML/min.} 
Then, CO gas was admitted in the chamber \rm{at a constant pressure}, and the temperature of the \rm{substrate was gradually decreased from 80 K to 8 K at a 
constant rate of 0.5 K/min}.  
Once the CO ice was deposited on top of the water ice, the substrate was warmed up from 8 K at a rate of 0.5 K/min, leading to the 
desorption of the CO molecules that were detected by a quadrupole mass spectrometer (QMS).

\subsection{Experimental results}
\begin{figure*} 
    \centering
    \includegraphics[width=0.55\textwidth]{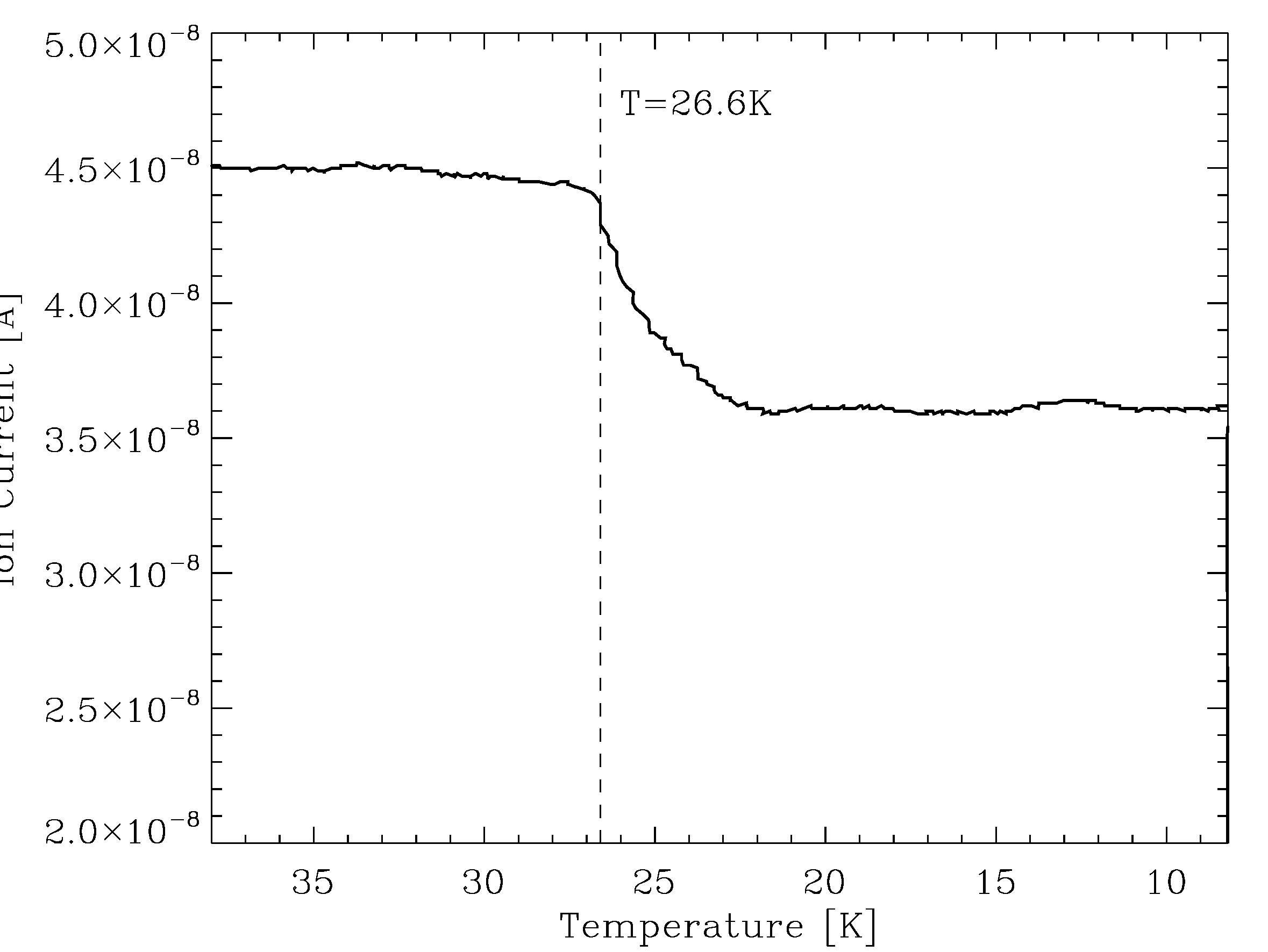}
\caption{The ion current of CO molecules in the gas phase measured using the QMS during cool-down of the substrate covered with water ice. The drop at 26.5~K is due to multilayer accretion.}
    \label{acc}
\end{figure*}

\begin{figure*} 
    \centering
    \includegraphics[width=0.45\textwidth]{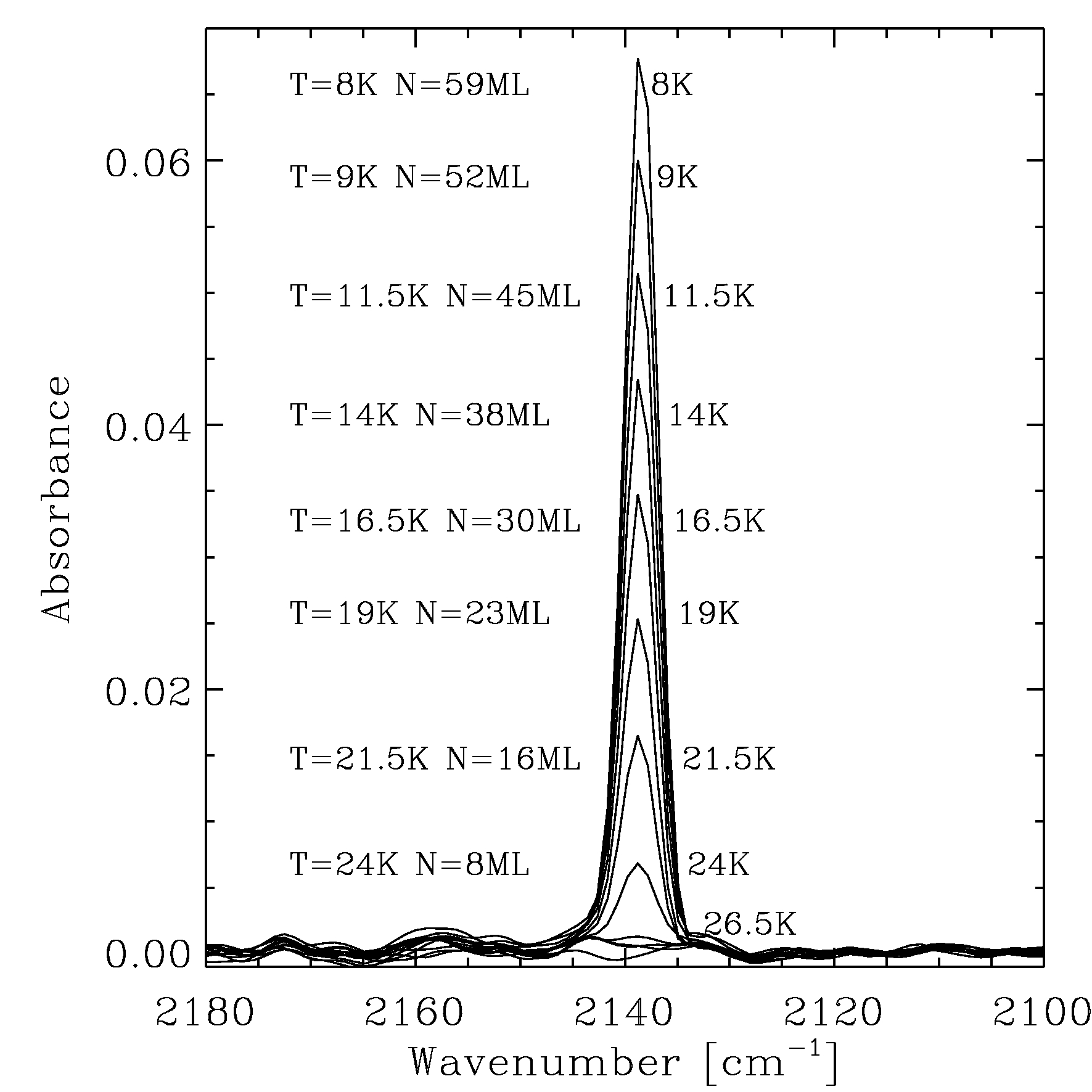}
    \includegraphics[width=0.45\textwidth]{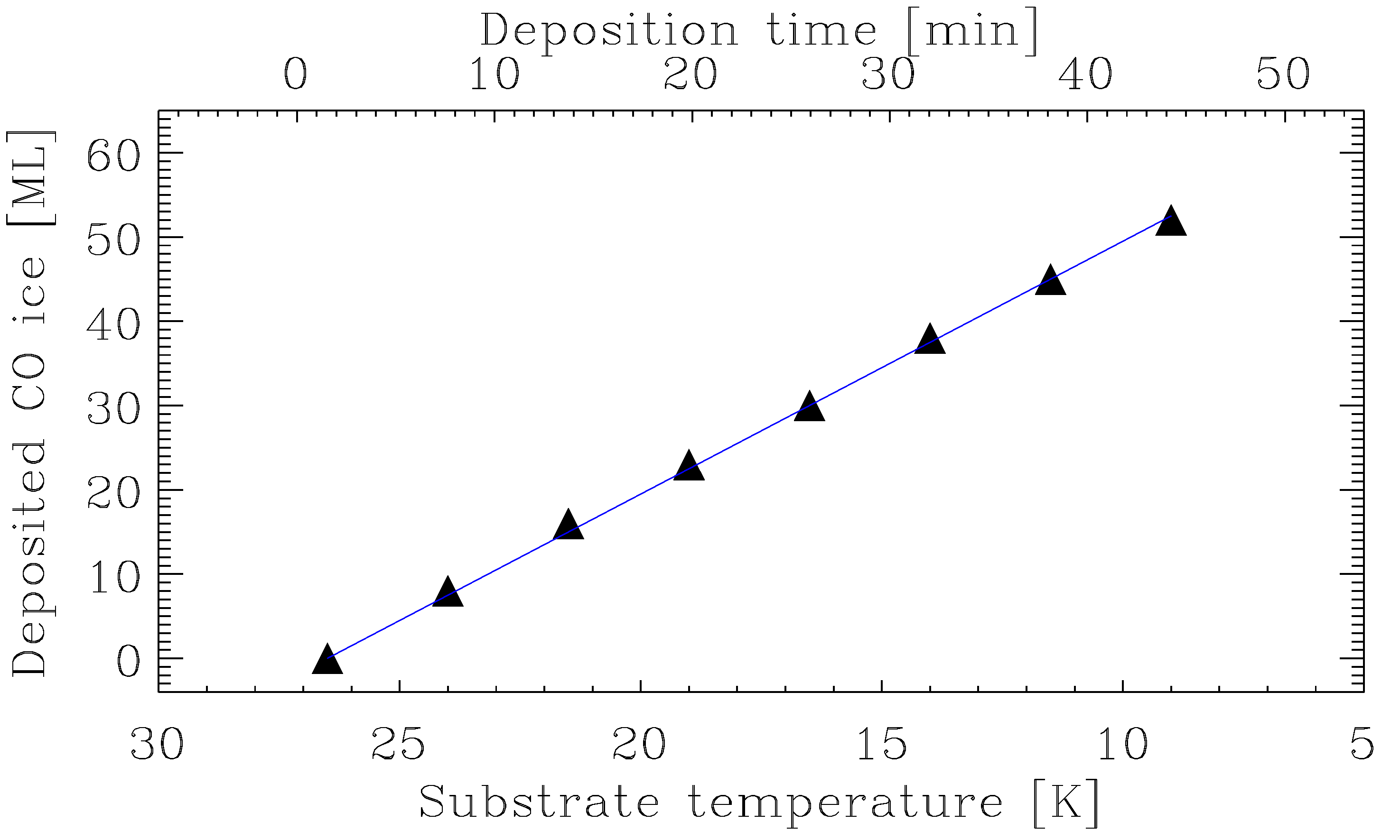}
    \caption{Left panel: number of accreted monolayers of CO ice measured with transmittance FTIR spectroscopy during the cool-down. 
    \rm{Right panel: integrated CO ice column density as a function of the temperature. Since the cooling rate was constant, 
    a linear increase in the deposited ice column density with temperature means a constant accretion rate for this temperature range.}}
    \label{acc2}
\end{figure*}

Figure~\ref{acc} shows the CO \rm{in the gas phase as measured by} the QMS \rm{while} the temperature of the substrate was cooled down from 80 to 8~K with a rate of 0.5~K/min. 
\rm{
The drastic decrease observed at 26.5 K is due to accretion of CO molecules on the substrate.  
This is confirmed by IR spectroscopy, as shown in Fig. \ref{acc2}, left panel. 
The CO IR band at $\sim$2139 cm$^{-1}$ was not observed at temperatures higher than 26.5 K. 
This means that only 1 ML(at most) of CO (which is the sensitivity of our FTIR spectrometer) 
could have been accreted on top of the ASW  before that temperature was reached. 
However, once the temperature decreases below 26.5 K, the solid CO IR feature is observed, 
increasing its intensity at a constant rate (see Fig. \ref{acc2}, right panel), which corresponds to an accretion rate of 1.5 ML/min. 
This feature does not present any shoulder at $\sim$2152 cm$^{-1}$, typical of CO molecules interacting with dangling 
OH bonds (see, e.g., \citealt{collings2003b}; \citealt{martindomenech2014}). Therefore, CO diffusion into the ASW structure does 
not take place in our experiment, at least to a significant extent. 
This is due to the lower porosity of the water ice deposited at  
80 K compared to that deposited at lower temperatures (see, e.g., \citealt{bossa2012}). 
This shoulder at $\sim$2152 cm$^{-1}$ may not be observed either in astronomical spectra (see, e.g., \citealt{cuppen2011}).}  

\begin{figure*}
    \centering
    \includegraphics[width=0.55\textwidth]{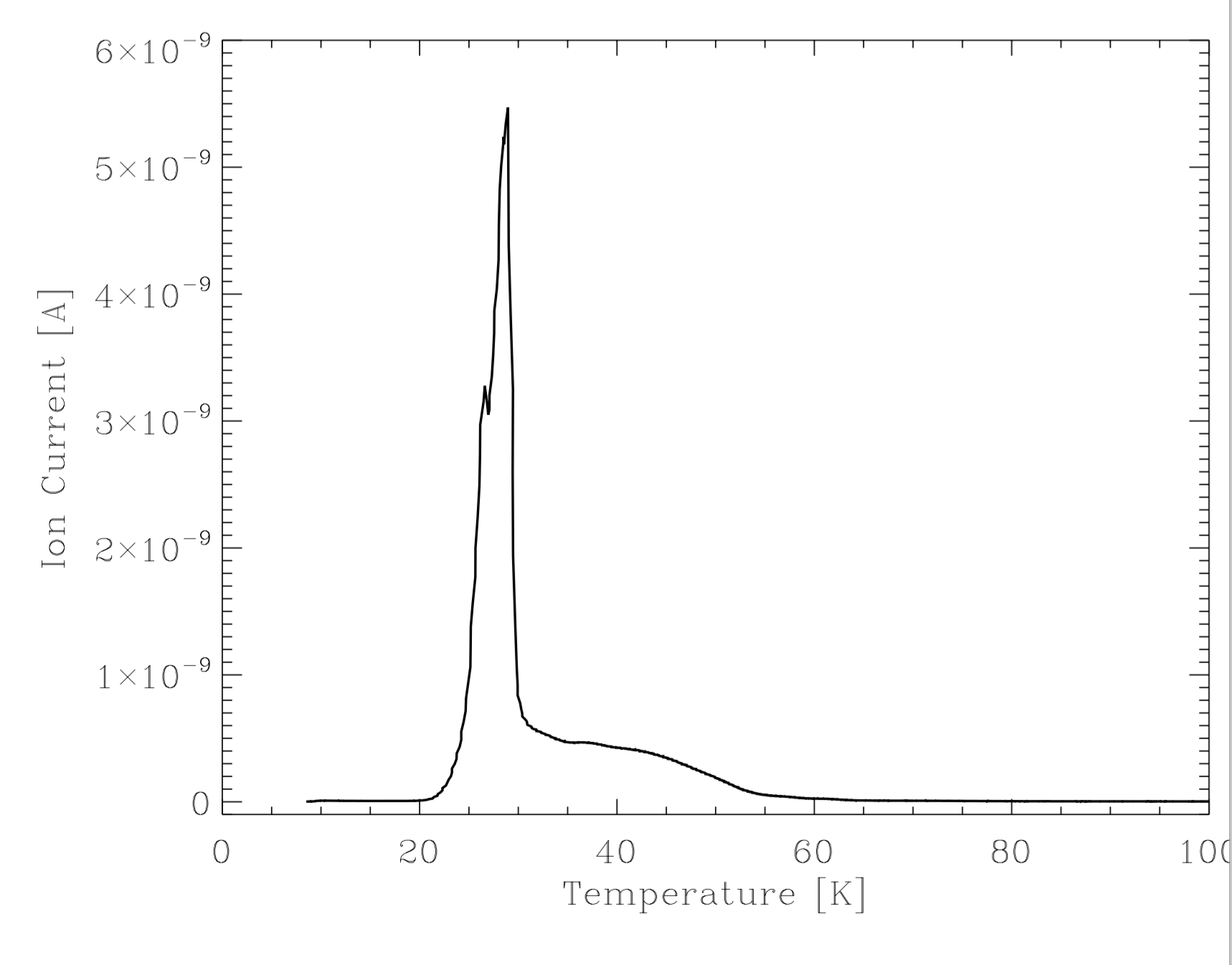}
    \caption{TPD experiment using the QMS corresponding to the CO ice accreted during cool-down of the substrate covered with water.}
    \label{tpd2}
\end{figure*}

\rm{Once the substrate reached a temperature of 8 K, and the temperature increased at a rate of 0.5~K/min (TPD). The desorbing CO molecules were detected by the QMS (see Fig.\ref{tpd2}). While an important desorption peak can be seen near 30~K, corresponding to the desorption of the bulk of the CO molecules, an extended shoulder ranges from 30~K to $\sim$60~K, corresponding to the desorption in the sub-monolayer regime (\citealt{noble2011}).}

\rm{This TPD behavior is similar to what was previously reported for CO ices accreted at a fixed temperature under different conditions (see, e.g., \citealt{collings2004};\citealt{martindomenech2014}). Therefore, at first glance, the deposition conditions do not seem to affect the subsequent thermal desorption process. } 


\subsection{Experimental setup: the Interstellar Photoprocess System}
\begin{figure*} 
    \centering
    \includegraphics[width=8cm]{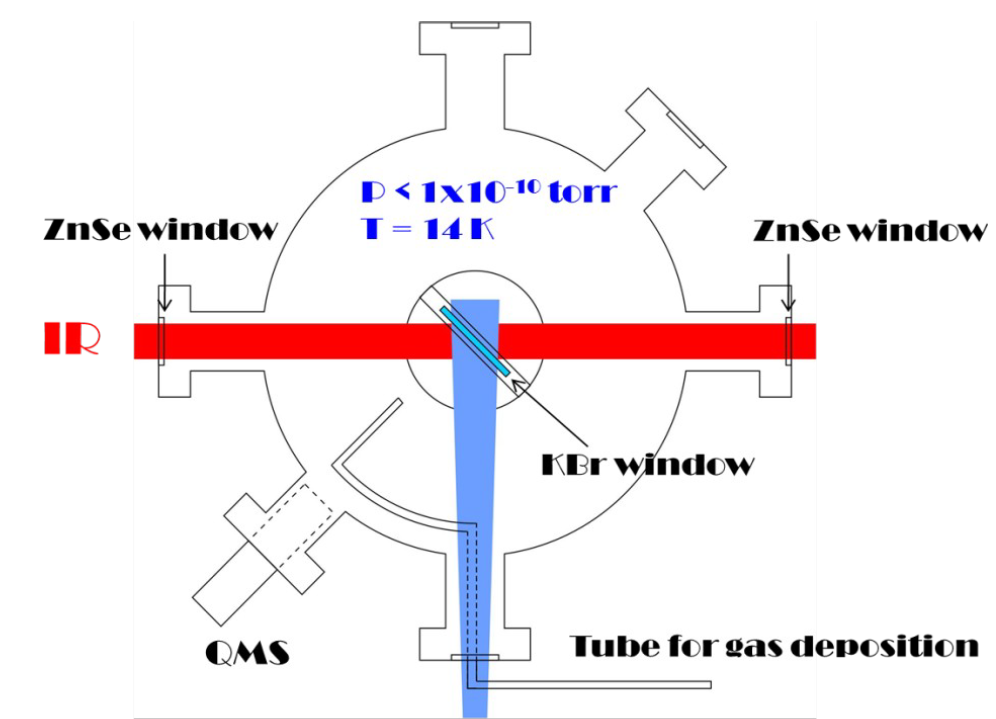}
    \caption{Schematic representation of the Interstellar Photoprocess System.}
    \label{ips}
\end{figure*}

Additional experimental simulations 
corresponding to the CO ice accretion at a constant temperature of 14 K, and subsequent TPD,    
were performed using the Interstellar Photoprocess System (IPS) described in Chen et al. (2014). 
\rm{The aim of these experiments was to confirm that a CO ice accreted at different conditions from that used in the experimental simulations 
described in the previous section presented similar TPD curves. In addition, as mentioned in Section 1, laser interference and infrared spectroscopy used in these series of experiments in order to get a better understanding of the desorption process.}  
A schematic representation of IPS is shown in figure~\ref{ips}. 
IPS is an ultra-high-vacuum (UHV) chamber with a base pressure of 1.3 $\times$ 10$^{-10}$ mbar. 
\rm{The substrate for interstellar solid analogs (usually a KBr window)} is located at the sample holder, 
placed on the tip of a cold finger from a closed-cycle helium cryostat (CTI-M350),   
which reaches temperatures as low as 14 K. 
Two silicon diodes are used to monitor the temperature of both the substrate and the cold finger, with an accuracy of 0.1 K. 
\rm{As in the ISAC setup, the species} were allowed to enter the UHV chamber through a capillary tube, 
condensing onto a KBr substrate \rm{and forming the ice analogs}. 
A FTIR spectrometer (model ABB FTLA-2000-104) equipped with a 
mercury-cadmium-telluride (MCT, more sensitive than a DTGS) detector monitors the solid sample. 
\rm{In this case,} the ice sample position is at 45 $\deg$ from the IR beam (see Fig. \ref{ips}) 
and, therefore, the thickness experienced by the IR beam (effective ice thickness) is larger than the actual ice thickness by a 
factor of $\frac{\sqrt{2}}{2}$. This is taken into account during the experimental simulations.  
A QMS covering the range of 1 - 200 amu with 0.5 amu resolution, provides monitoring of the 
introduced gas during the deposition, 
and measures the presence of desorbing molecules in the gas phase during the warm-up phase. 

\rm{In addition, a laser reflective interference system (not included in the ISAC setup)  
counts with a He-Ne laser ($\lambda$ = 632.8 nm), and a 
power meter (Newport M835) calibrated for measuring the reflected intensity of the laser at 632.8 nm (power $<$ 5 mW). 
The He-Ne laser reaches the solid sample with an incident angle of 2 degrees with respect to the normal, 
and the reflective light is subsequently directed by a mirror to the power meter (see Fig. \ref{ips}). 
The reflected intensity of the laser oscillates between constructive and destructive interference, leading to a sinusoidal pattern. 
This interference pattern was monitored during both a blank experiment (warming up of a KBr substrate alone)  
and the TPD of a CO ice deposited directly onto the bare KBr substrate at 14 K.   
During the blank experiment, the sinusoidal interference pattern responded to the (constructive and destructive) 
interference between the light reflected from the front KBr surface and the rear KBr surface.  
When the KBr window is held at a specific temperature, its thickness is constant and does not cause 
any variation of light interference, but this changes during the warm up, leading to a variation in the sinusoidal pattern.  
On the other hand, during the TPD of the deposited CO ice, the pattern responded to the interference between the light reflected 
from the ice surface and the KBr window, and changed as the CO molecules desorbed from the ice.}  

\subsection{Results}

\begin{figure*} 
    \centering
    \includegraphics[width=0.55\textwidth]{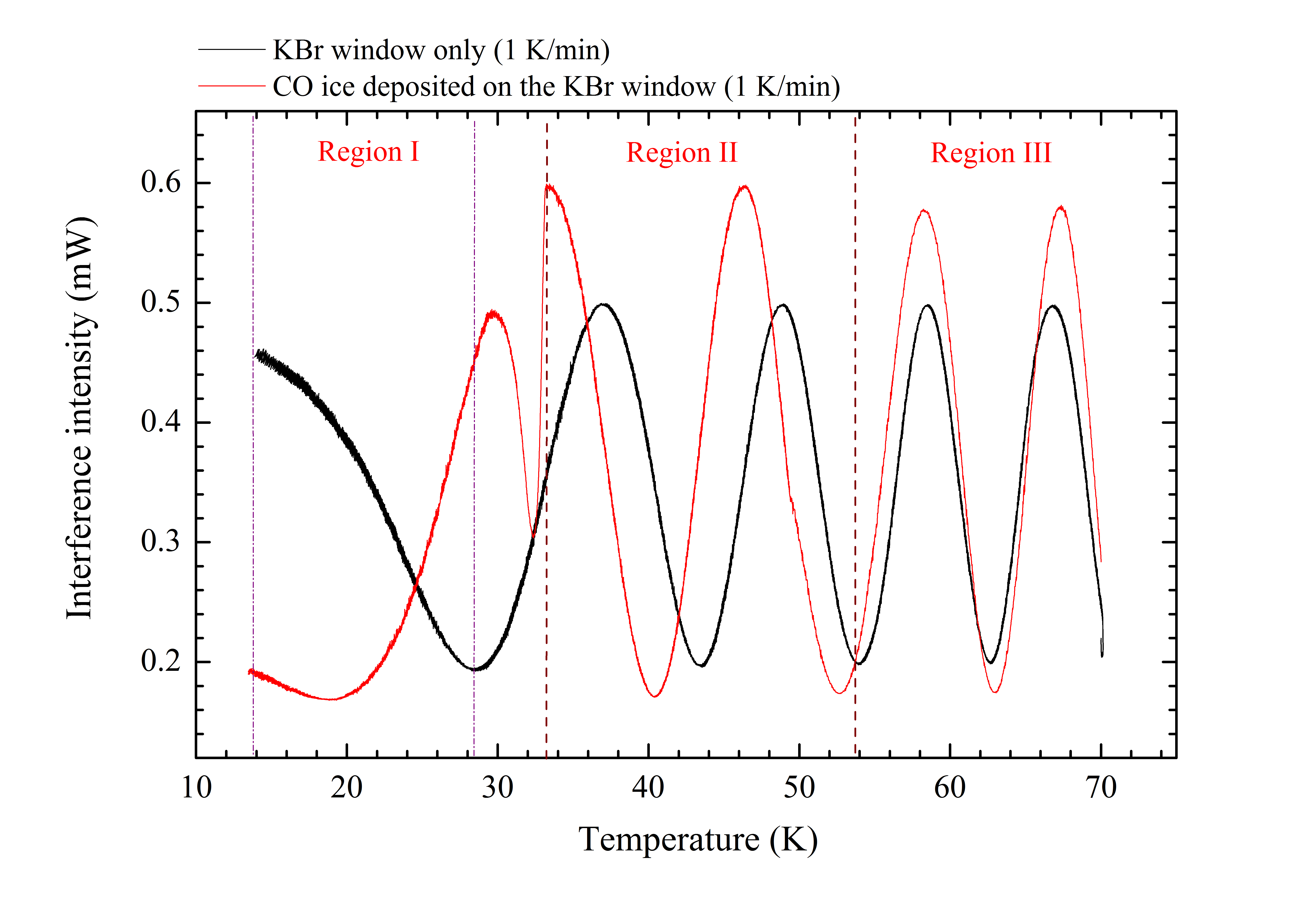}
    \caption{TPD experiment using laser interference to study the solid CO sample.}
    \label{ir}
\end{figure*}

\begin{figure*}[h]
    \centering
    \includegraphics[width=0.45\textwidth]{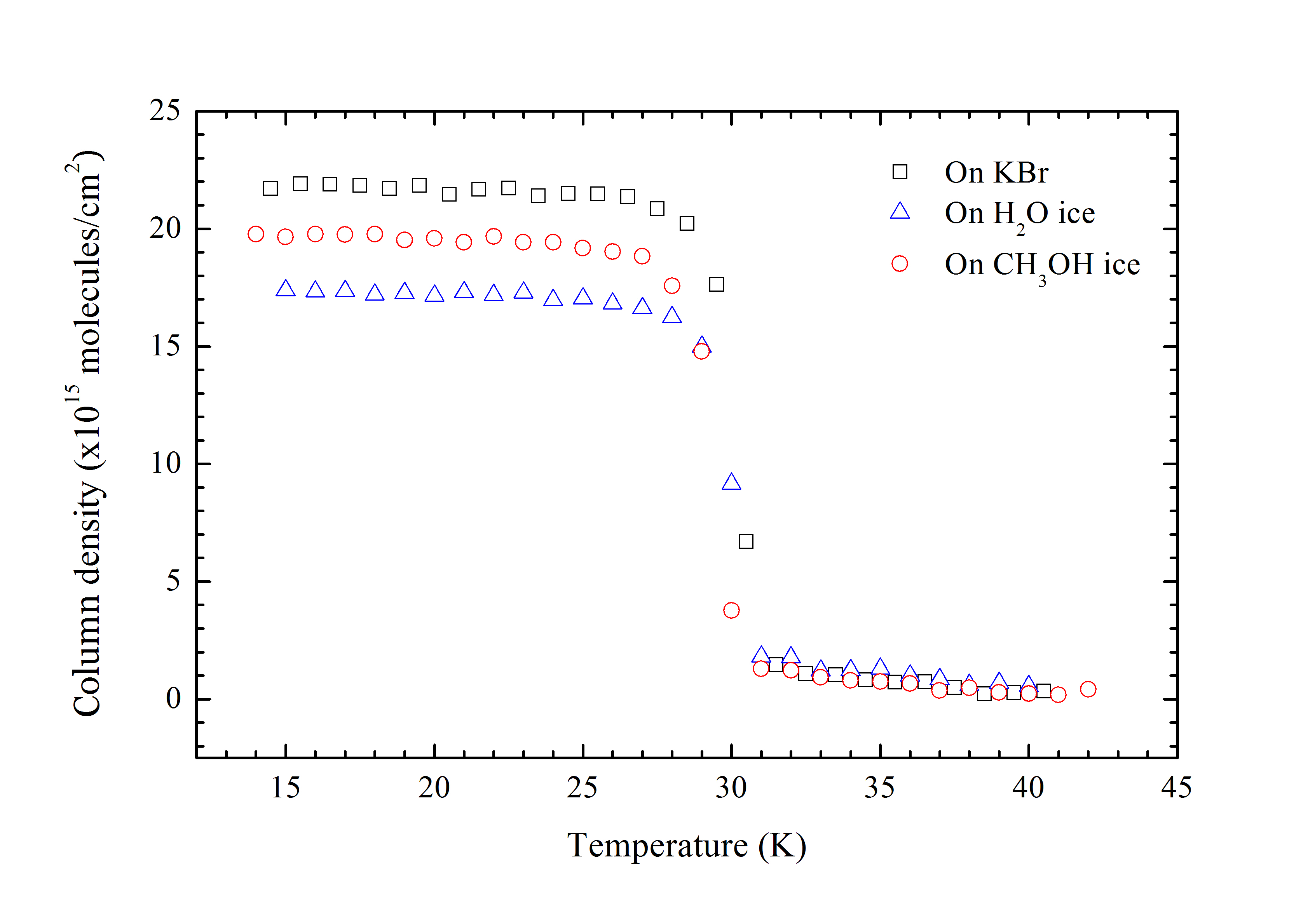}
    \includegraphics[width=0.45\textwidth]{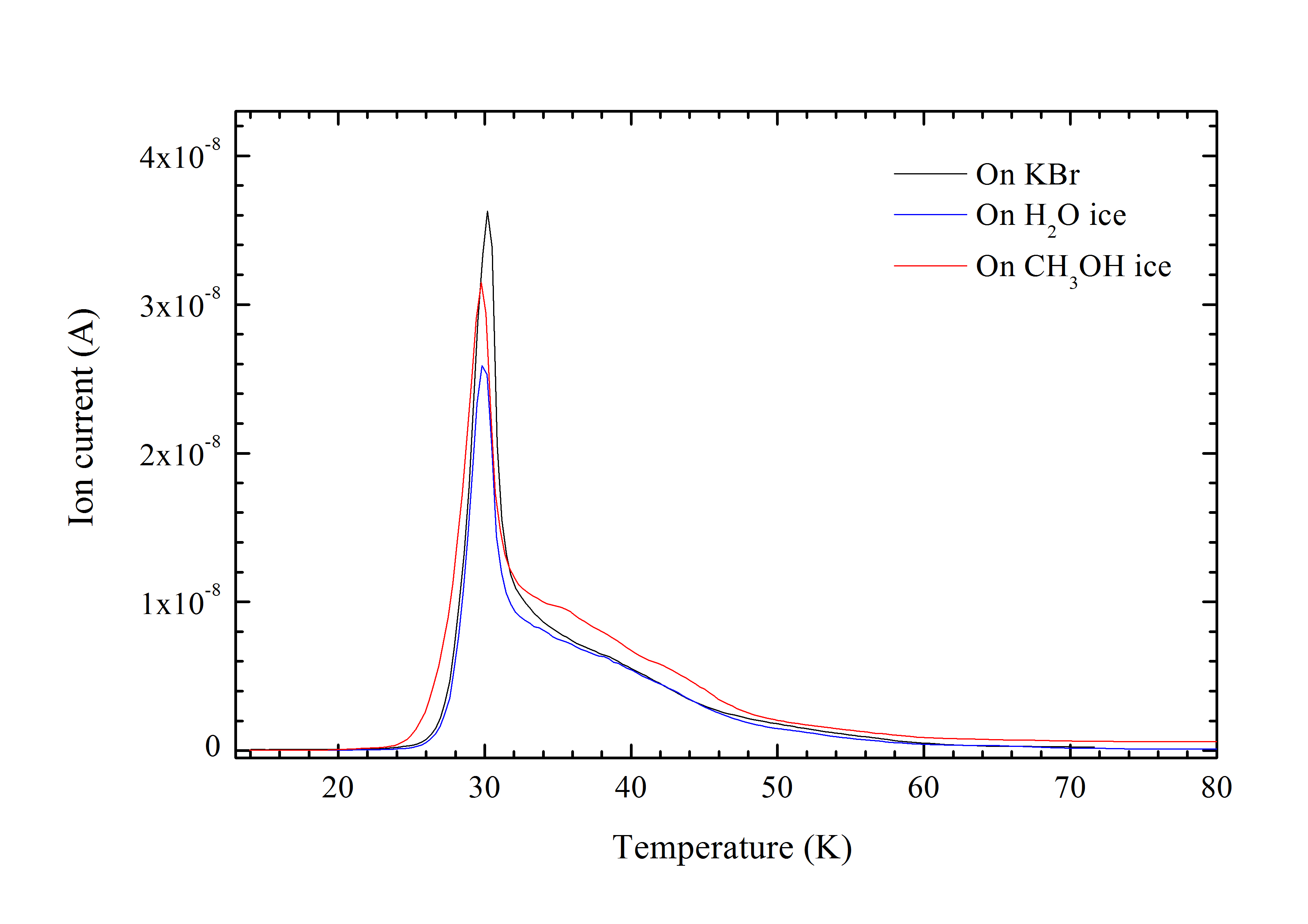}
    \caption{Left: IR measurement of the ice during the TPD experiment, showing the number of monolayers on the surface (1 ML=10$^{15}$ molecules/cm$^2$). Right: In these experiments, the QMS was used to record the CO molecules in the gas phase.}
    \label{tpdqms}
\end{figure*}

\rm{The evolution of the CO ice was studied by using both laser interference to follow the CO solid sample (Fig. \ref{ir}, and Fig.\ref{tpdqms} left panel) and the QMS to follow the TPD and therefore the CO in the gas (Fig. \ref{tpdqms}, right panel). In this experiment, the sample temperature was increased from 14 K to 70 K at an identical rate of 1 K/min. }

The results corresponding to \rm{the blank experiment (warming up of a KBr substrate alone) and the TPD of a CO ice deposited onto a bare KBr substrate at a constant temperature of 14 K using laser interference to study the subsequent desorption of the CO ice  are shown in Fig. \ref{ir}. 
The interference pattern 
as a function of the temperature during TPD of the CO ice is shown in red, and compared to the blank experiment in black. 
The variation of the sinusoidal pattern during the TPD experiment is due to the thickness variation 
of both the KBr substrate and the CO ice. 
Therefore, the laser interference pattern cannot be used to estimate the thickness variation of the CO ice alone
because it is difficult to determine the contributed weights from the ice sample and the substrate to this effect. 
However, the differences in the variation of this sinusoidal pattern during the TPD of the CO ice with respect to the blank 
experiment allow us to determine if some CO is retained on the KBr surface, and if its thickness considerably changes. 
Between 13 K and 28 K, the measurements indicate that no important changes of the ice thickness are observed for temperatures prior to CO desorption 
(region I), since the variation of the sinusoidal pattern is similar to that of the blank experiment. 
Around 30K, a fast decrease and a subsequent increase of the laser intensity (red line) is observed due to the sublimation of CO 
molecules and the subsequent thickness variation. 
For temperatures higher than 32~K, in region II, the phase of the red line (CO ice) is delayed compared to the black line (blank 
experiment). 
This implies that some CO molecules are still present on the surface. 
In region III, above 54 K, the phases of the red line (CO ice) and black line (KBr substrate) become the same, which suggests that all CO molecules have desorbed.
The decrease of the CO ice column density due to thermal desorption was simultaneously monitored from integration of the IR band at $\sim$2139 cm$^{-1}$ during the TPD experiment, and is reported in the left panel of Fig. \ref{tpdqms}. 
The IR measurements support the idea that approximately 1 monolayer of CO remains on the surface at temperatures above $\sim$30 K (region II in Fig. \ref{ir}) and desorbs at T$\le$60 K.}

The TPD measurements are reported in the right panel of Fig. \ref{tpdqms}.
CO deposition was studied on three different types of surfaces, namely: the aforementioned bare KBr substrate (this experiment is represented in black in Fig. \ref{tpdqms}), 
water ice previously deposited at 80 K with a rate of 0.1 ML/s (blue), and methanol ice deposited at the same temperature with a rate of 0.07 ML/s (red). 
Results are similar in the three cases, and also to those presented in the previous section, thus confirming that the deposition conditions do not greatly affect, at first glance, the subsequent thermal desorption process.
CO ices stay adsorbed on the substrate until $\sim$ 30~K. At this temperature, the CO molecules that constitute the bulk of the ice desorb into the gas phase and are detected by the 
QMS, leading to the desorption peak in the right panel of Fig. \ref{tpdqms}. 
From $\sim$30~K to $\sim$50~K, the TPD measurements from the QMS show a slow decrease of the signal. 
This, along with the IR and laser interference measurements (only shown for the TPD of CO ice deposited onto the KBr substrate in Fig. \ref{ir}), suggests that around 1 ML of CO is still present on the different surfaces until $\sim$54~K, as expected from previous experiments (see, e.g., \citealt{noble2011}).

The slow decrease in intensity of the TPD signal inferred from the QMS data collected during the three experiments suggests that 
\rm{these} CO molecules are strongly bound to the different surfaces, with binding energies ranging between $\sim$1000 K and 1600~K. 
The binding energies of the molecules can be directly estimated from the maximum desorption temperature as described in  \cite{luna2017}, where E$_{\rm{bin}}$ (in K)=30.9 $\times$ T$_{\rm{des}}$ (in K). 

\subsection{Accretion of CO: summary of experimental results}
The two experiments differed in the CO ice deposition temperature (decreasing temperature from 80~K to 8~K \rm{versus} fixed at 14~K), 
as well as in the resulting deposition rates (1.5 ML/min \rm{versus} 7.3 ML/min, respectively). 
Deposition of CO with decreasing surface temperature in the ISAC setup shows that CO accretes on the surface \rm{at temperatures below} 26.5~K. 
At this temperature, the drop in the CO ion current, due to the important accretion on the substrate, reflects the multilayer regime, where the temperature is low enough so that CO can be bound to CO ice.  This suggests that CO molecules deposited at 26.5~K are able to find binding sites on the CO ice with binding energies of the order of 800~K. 
\rm{The accretion rate was found to remain constant as the substrate temperature decreased, suggesting that the sticking coefficient of the CO molecules remains constant with temperature, as reported in \citealt{dawson1965}. }
In addition, the two experiments show that deposition at the two different conditions mentioned above 
result in similar TPD curves (\rm{TPD curves are also very similar when the CO ice is accreted on different surfaces, as shown in Fig. \ref{tpdqms}}), in which CO desorbs at $\sim$30~K in the multilayer regime, while the sub-monolayer regime desorbs between 30 K and 60~K (see also \citealt{noble2011}). 
\rm{The QMS measurements were confirmed by IR spectroscopy and reflective laser interference of the solid sample in the case of experiment 2 performed with the IPS setup. 
This seems to indicate that no structural changes in the CO ice during warming up could be detected during our experiments, which is in agreement with previous work (see, e.g., \citealt{munozcaro2016}). However, a phase change from amorphous to crystalline CO ice is known to take place prior to thermal desorption (\citealt{kouchi1990}). This indicates that the transition from amorphous CO as deposited at low temperatures (typically between 8 and 20~K) to crystalline CO must occur at temperatures close to the peak of desorption in the TPD curve see section 2.6}

\rm{A desorption peak at around 50~K attributed to monolayer desorption has been reported in some previous experimental studies (see, e.g., \citealt{collings2004}) but was not observed in the TPD curves shown in the previous sections. When the TPD is performed with a heater located close to the sample, a temperature gradient may be created from the sample to other parts of the cryostat. In this scenario, desorption of CO molecules adsorbed on surfaces different from the substrate can take place when the temperature of the substrate is higher than the temperature of this surface, leading to a desorption peak in the TPD curve that is actually an artifact. This is not the case of the present work.}


\rm{While the TPD for CO deposition at fixed T=14~K or at 80~K downwards show similar features, their representation in logarithm scale shows a small bump near 20~K. This bump is most prominent for CO deposited at T=14~K (red) than at 80~K downwards (black). This indicates that CO molecules with lower binding energies are present on the CO ice in both cases, but also that deposition at 14~K provides more of such weakly bound CO molecules compared to deposition from 80~K downwards. }\rm

\begin{figure*}[h]
    \centering
    \includegraphics[width=0.45\textwidth]{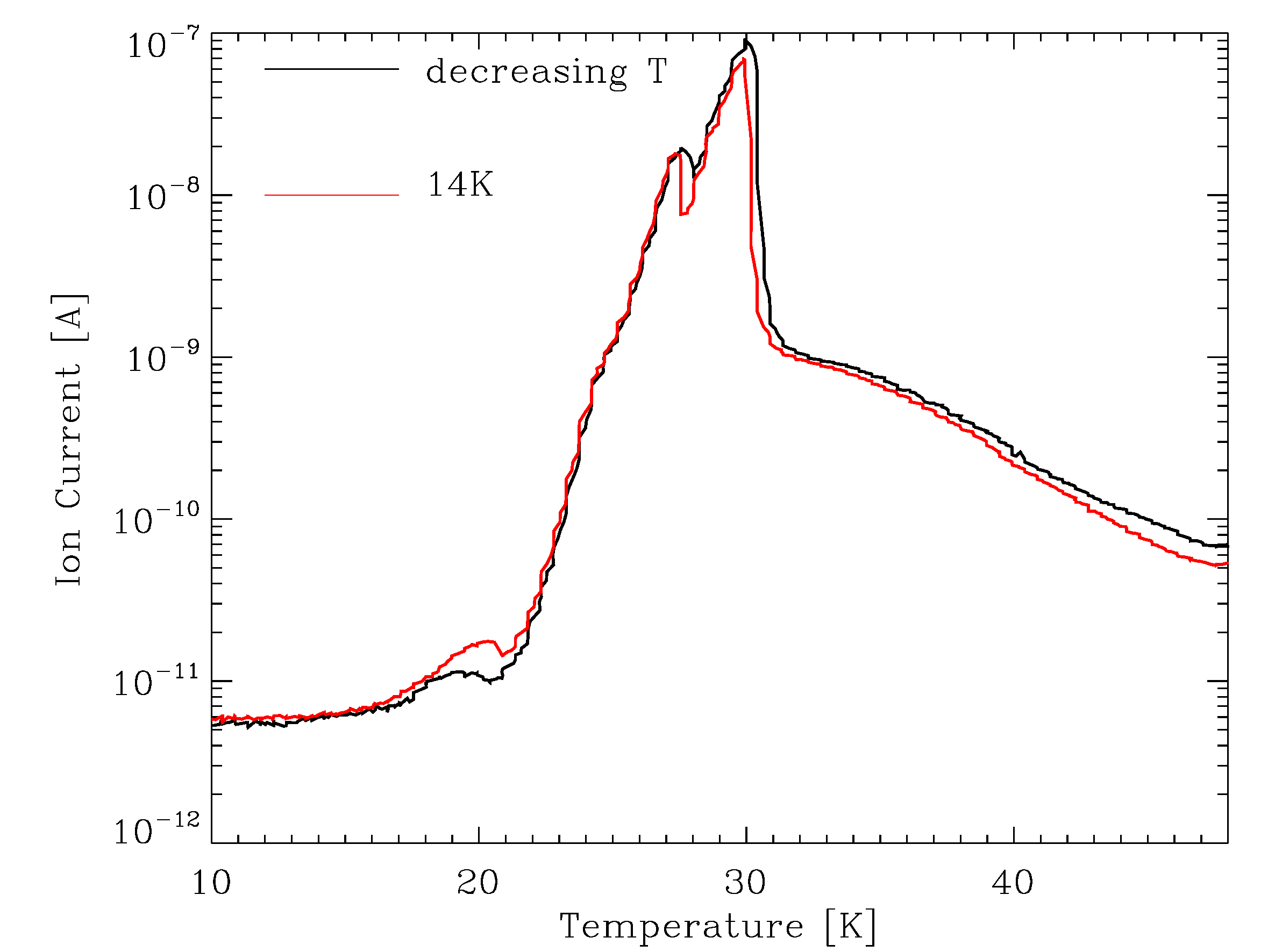}
    \caption{TPD in logarithmic scale showing the desorption of CO molecules from CO ice deposited at 14~K (red) or deposited from 80~K downwards (black).}
    \label{tpdlog}
\end{figure*}

\subsection{Amorphous versus crystalline CO ice.}
\begin{figure*}[h]
    \centering
    \includegraphics[width=0.45\textwidth]{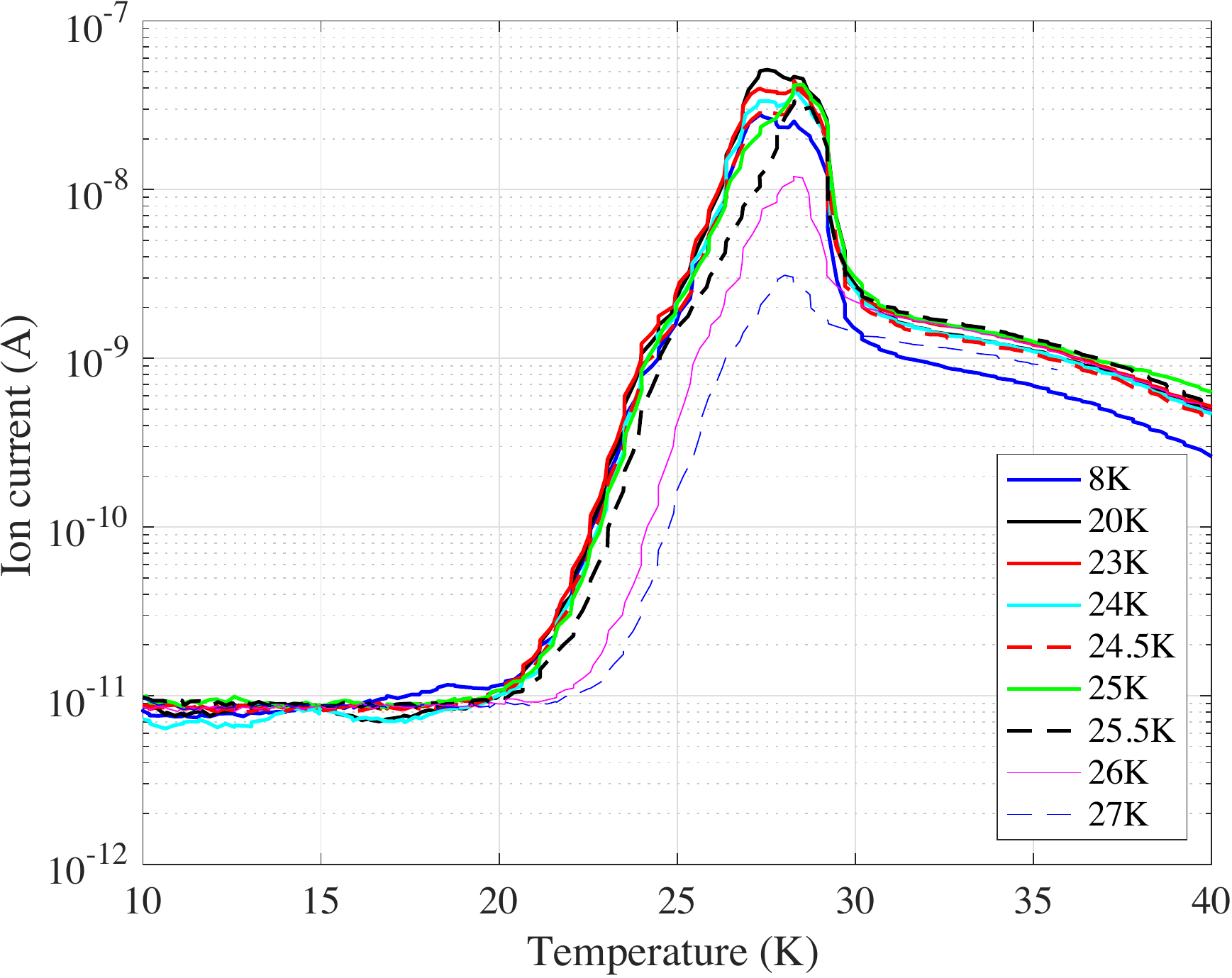}
    \caption{TPD in logarithmic scale showing the desorption of CO molecules from CO ice deposited at fixed temperatures from 8~K to 27~K.}
    \label{amorph_crys}
\end{figure*}

\rm{As mentioned in the previous section, the transition between amorphous to crystalline CO ice occurs at temperatures close to the peak of desorption in the TPD curve. We performed a series of TPD measurements for CO deposited at temperatures from 8~K to 27~K. For each experiment of CO deposited at fixed temperature, the temperature is then decreased to 8~K and the TPD is then performed until 50~K with a rate of 1~K/min. Figure \ref{amorph_crys} shows the different TPD peaks for the different deposition temperatures. The first obvious result is that TPD for low temperatures deposition is composed of 2 peaks, one located at 27~K and the other at 28~K.  For the lowest deposition temperatures, at 8 and 20~K, the peak at 27~K is more intense, while for deposition at 23 and 24~K, the peak at 28~K is dominating. For deposition at 25K, although the beginning of the TPD is the same as for lower temperatures, only one peak is seen. This is also the case for deposition at  25.5~K and 26~K, which also show an increasing displacement at the beginning of the TPD. The change in the intensity of the peaks indicates that the phase change occurs above 24K (and is already evident at 25K), but the stronger shifts at 26 and 27 K indicate that there is still a fraction of amorphous ice mixed with the crystalline phase. Our study is in agreement with previous studies showing the phase change of CO from amorphous to crystalline being close to the desorption temperature (\citealt{kouchi1990}).}\rm

\section{Simulations}
We used a step-by-step Monte Carlo simulation to follow the formation of CO ices through deposition and \rm{subsequent} evaporation in the gas phase. Our model is described in \cite{cazaux2016}. CO molecules originating from the gas phase arrive at a random time and location \rm{to the substrate}, and follow a random path within the ice. The arrival time depends on the rate at which gas species collide with the surface (section 3.1). The molecules arriving on the surface can be bound to the substrate and to other CO molecules through van der Waals interactions. The binding energy of each CO molecule depends on its number of neighbours, as described in section 3.2. \rm{Our theoretical approach to estimate binding energies is similar than the one from \cite{cuppen2007} and \cite{garrod2013} but here we had to determine how the binding energy of a single CO molecule increases with the number of CO neighbours around in order to reproduce the experimental results.}\rm\ Depending on its binding energy, the CO molecules diffuse on the surface/in the CO ices. The diffusion is described in section 3.3. During warming-up, the CO molecules can evaporate from the substrate/ices (section 3.4). The number of molecules evaporating as a function of the temperature, corresponds to the QMS measurements.

\subsection{Accretion}
In our model, we defined the surface as a grid with a size of 20$\times$20 sites. CO molecules from the gas-phase arrive on the grid and can be bound to the substrate (that we choose as water substrate to mimic the experiments) and/or to adsorbed CO molecules through van der Waals interactions. The accretion rate (in s$^{-1}$) depends on the density of the species, their velocity, and the cross section of  the surface, and can be written as:
\begin{equation}
R_{\rm{acc}} = n_{\rm{CO}} v_{\rm{CO}}  \sigma  \rm{S}, 
\end{equation}
$v_{\rm{CO}}=\sqrt{8  k  T_{\rm{gas}}/(\pi m_{\rm{CO}})} \sim 2.75 \times 10^4 \sqrt{\frac{T_{\rm{gas}}}{100}}$ \rm{cm~s$^{-1}$ is the thermal velocity, S the sticking coefficient that we consider to be unity in this study. The cross section of the surface, $\sigma$, directly scales with the size of the grid we use for the simulations, which is 20$\times$20 sites in our calculations. Since the distance between two sites is 3 \AA, the density of sites is what is typically assumed, i.e. $\sim$(3 \AA)$^{-2}$$\sim$10$^{15}$cm$^{-2}$. The cross section scales with the size of the grid considered in our calculations, as $\sigma$ $\sim$ (3$\times$ 10$^{-8}$$ \times$ 20)$^2$cm$^2$=3.6\ 10$^{-13}$ cm$^2$ . The deposition rate is therefore: $R_{\rm{acc}} = 1.7 \ 10^{-8}$ n$_{\rm{CO}}$ s$^{-1}$, for T$_{\rm{gas}}$=300~K. In order to mimic experimental conditions with deposition rates of 7.8 ML/min $\sim$ 52 molecules/s and 1.5 ML/min $\sim$ 10 molecules/s, we set the density of CO molecules in the gas in cm$^{-3}$ as being n$_{\rm{CO}}$= 3 $\times$ 10$^{9}$ cm$^{-3}$ and n$_{\rm{CO}}$=6 $\times$ 10$^{8}$ cm$^{-3}$ cm$^{-3}$ respectively (1 ML corresponds to 400 molecules on a 20 $\times$ 20 grid). 

\subsection{Building CO ices.}
The desorption of a CO molecule on a water ice surface is observed between 30-50~K which corresponds to binding energies ranging between 900 and 1500 (\citealt{he2016},\citealt{martin2014}, \citealt{noble2011}). To account for the fact that the water substrate is not homogeneous, we describe the initial surface of water ice with a random distribution of binding energies centered around 1200~K with a dispersion of 180~K as shown in \cite{noble2011}. \rm{In this sense, by using such a distribution, we are mimicking a non-flat and smooth substrate, as shown in many previous studies.}\rm\ If the molecules are not bound to the surface, but start to pile up, the binding energies of CO molecules, due to CO-CO interactions, increase with the number of surrounding neighbours. The lowest interaction between two CO neighbour molecules is around 16 meV (185~K; \citealt{karssemeijer2014}). In a multilayer regime, the binding energy of a CO molecule is about 830~K (\citealt{luna2014}, \citealt{noble2011}, \citealt{munozcaro2010}, \citealt{pontoppidan2006}, \citealt{Acharyya2007}, \citealt{collings2003a}). The binding energy as function of coverage of CO on ASW water ice has been reported by \cite{karssemeijer2014}. At low coverages, the binding energy of CO on the surface is $\sim$ 125 meV (1450~K), while for high coverages the surface becomes covered by CO molecules and the binding becomes $\sim$ 75 meV (870~K; TPD desorption temperature of 29~K, table 3 from \cite{martin2014}).

In order to estimate the binding energy of CO molecules as function of the number of neighbours, we use a simple approximation that is shown in Fig.~\ref{cluster}. The points show the interaction of a CO molecule with one single CO molecule, which is about 185~K and becomes of 860~K in the multilayer regime, which corresponds to 1 (CO adsorbed one CO molecule) to 3 direct neighbours (CO embedded on a top layer with 1 neighbour underneath and 2 neighbours around). By using a fit through these points we mimic a saturation for a high number of neighbours and can calculate the binding energy of a CO molecule as function of its number of neighbours nn:
\begin{equation}
E_{\rm{CO}}=-3360*(nn+1)^{-2}+1040
\end{equation}

In our calculations, we compute the binding energy by considering an effective number of neighbours nn, that scales with the distance between the neighbour CO$_{\rm{n}}$ and the CO molecule. In that sense, one neighbour would contribute as +1/(r$_{CO-CO_{\rm{n}}})^6$ depending on the CO-CO$_{\rm{n}}$ distance, to account for the fact that van der Waals interactions depend on distance as 1/r$^6$. For direct neighbours, this distance is 1, while for neighbours on the sides this distance is $\sqrt 2$ and for the neighbours located in a corner the distance is $\sqrt 3$. 

\begin{figure*} [h]
    \centering
    \includegraphics[width=0.53\textwidth]{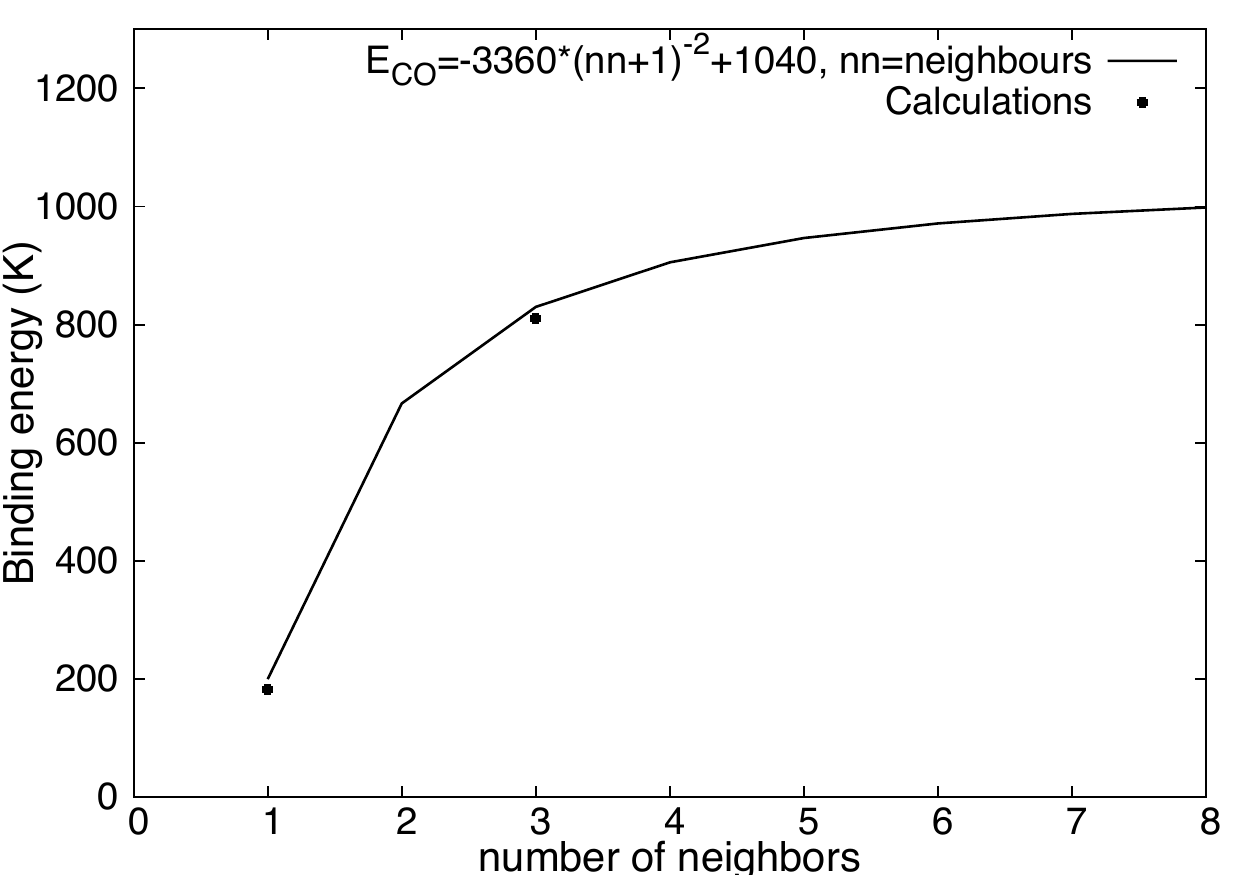}
    \caption{Binding energies of CO molecules as function of neighbours. The points correspond to the calculations from \cite{karssemeijer2014b}. }
    \label{cluster}
\end{figure*}

\rm{In the theoretical model from \cite{cuppen2007},  
the binding energy of individual molecules is the sum of the binding with their neighbours. \cite{garrod2013} performed 
off lattice KMC method to compute the reactivity and porosity of ices, also considering that the binding energy of one species 
is the sum of the pair-wise interaction potentials with its neighbours. While this method is more sophisticated than the present 
method since it allows to determine the distance of the species explicitly, we here consider the distance between CO molecules to 
be equal, and concentrate on defining the binding energies as function of neighbours. However, because our method does not 
compute the distance between CO molecules, the diffusion could be different because (1) the distances are not identical (2) 
using a Lennard-Jones potential, such as in \cite{garrod2013} instead of a square barrier, would change the diffusion efficiency. 
This could have an impact on the number of holes/micropores present in the ices deposited at very low temperatures.}\rm\

\subsection{Diffusion}
A recent study on the diffusion of CO on hexagonal water ice surface shows that diffusion barriers are of the order of 50~meV (\citealt{karssemeijer2014}), which represent only $\sim$ 30$\%$ of the binding energy. We define the diffusion rates depending on the number of neighbours interacting with the CO molecules, that we call $nn$ in the above section. For a position (i,j,k) of a CO molecule in the grid, we calculate the associated binding energy Ei and identify the possible sites where the molecule can diffuse to as i$\pm$1; j$\pm$1; k$\pm$1. The final binding energy Ef is calculated as function of the neighbours present around this site. The diffusion rate, from an initial site with an energy Ei to a final site with an energy Ef, is illustrated in Fig.\ref{barrier}. The barrier to go from Ei to Ef  is defined as follows if Ei$\le$Ef (Fig.\ref{barrier}; left panel): 
\begin{equation}
\rm{Ed= \alpha \times min(Ei,Ef)}, \hspace{1cm} if \  \rm{Ei<Ef}
\end{equation}
If Ei$>$Ef, on the other hand, the barrier becomes (Fig.\ref{barrier}; right panel)
\begin{equation}
\rm{Ed= \alpha \times min(Ei,Ef) +\Delta E}, \hspace{1cm} if \ \rm{Ei>Ef}
\end{equation}
with $\Delta$E = max(Ei,Ef)-min(Ei,Ef). By defining the barriers in such a manner, we do take into account microscopic reversability in this study (\citealt{cuppen2013}). The barriers to move from one place to another should be identical to the reverse barrier. The diffusion barriers scale with the binding energies with a parameter $\alpha$. \rm{While this parameter is found to be of 30$\%$ for CO on water ice from \cite{karssemeijer2014} which is of the same order of the water-on-water diffusion derived experimentally (\citealt{collings2003b}), recent studies also point out to very small values for the CO diffusion barriers (\citealt{lauck2015}). However, the CO-on-CO diffusion has not been determined, but studies highlight the large differences between bulk and surface diffusion (\citealt{ghesquiere2015}).\rm\ The diffusion parameter $\alpha$ sets the temperature at which CO molecules can re-arrange and diffuse in the ices to form more dense ices. In this study, we reproduce experimental results in order to constrain the diffusion parameter.}\rm

\begin{figure*} [h]
    \centering
    \includegraphics[width=0.5\textwidth]{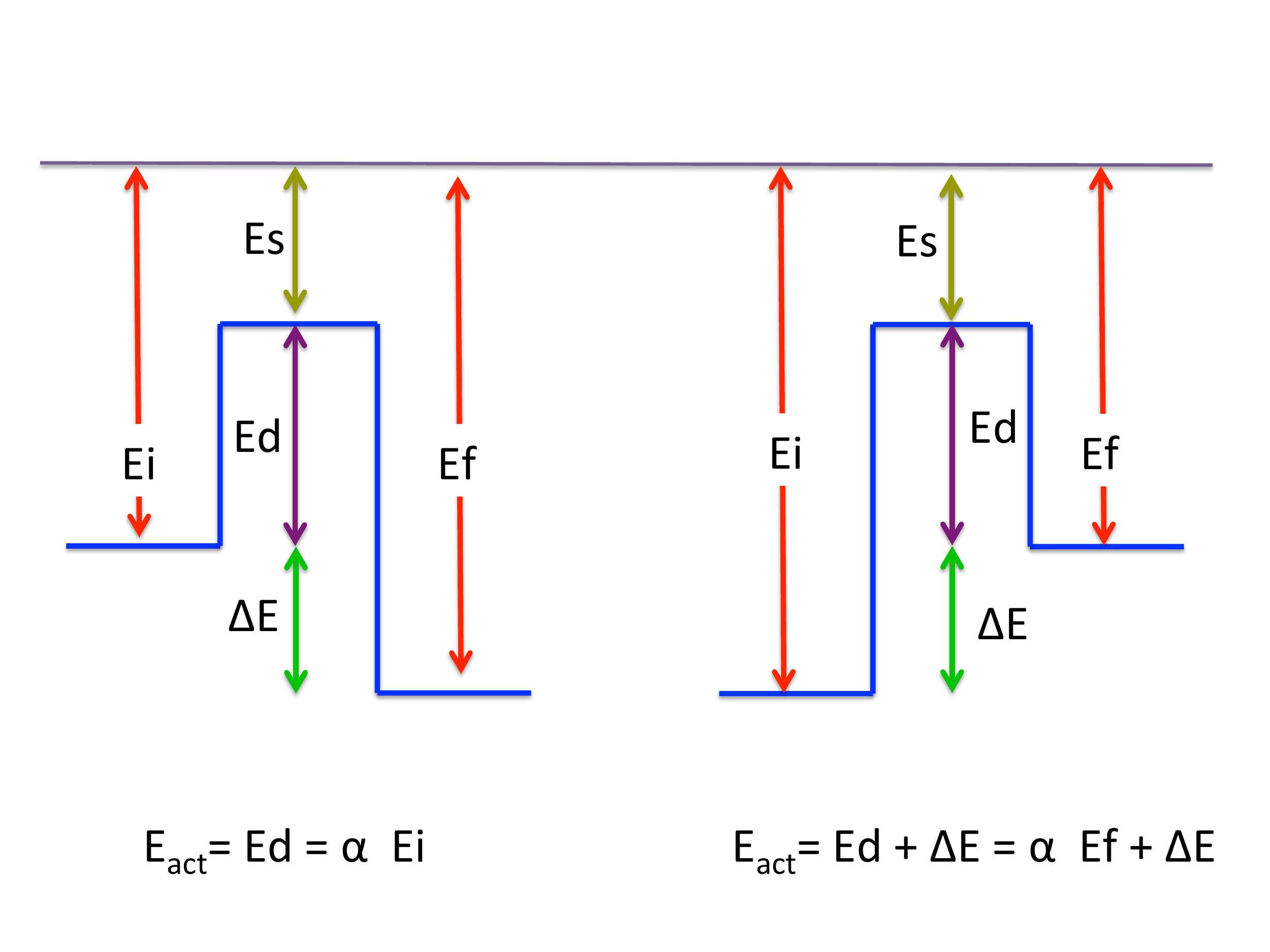}
    \caption{Diffusion barrier to go from a initial site with energy Ei to a final site with energy Ef for the case where Ei$<$Ef (left panel), and Ei$>$Ef (right panel).}
    \label{barrier}
\end{figure*}

The diffusion rate, in s$^{-1}$, for a CO molecule can be written as: 
\begin{equation}
R_{\rm{diff}} = 4 \times \sqrt{\frac{\rm{Ei-Es}}{\rm{Ef-Es}}}\times \nu \exp\left({-\frac{E_{\rm{act}}}{\rm{T}}}\right), 
\end{equation}
where $\nu$ is the vibrational frequency of a CO molecule in its site (that we consider as 10$^{12}$ s$^{-1}$), T is the temperature of the substrate (water ice or CO ice) and Es is the energy of the saddle point, which is Es=(1-$\alpha$)$\times$ min(Ei,Ef). This formula differs from typical thermal hopping because the energy of the initial and final sites are not identical (\citealt{cazaux2004}).

\subsection{Evaporation}
The CO molecules present on the surface can return into the gas phase because they evaporate. This evaporation rate depends on the binding energy of the species with the surface/ice. As mentioned previously, the binding energy of a CO molecule depends on its number of neighbours, or wether the molecule is directly bound to the surface. The binding energy Ei of the CO molecule sets the evaporation rate as:
\begin{equation}
 R_{\rm{evap}}(X)=\nu \exp\left({-\frac{Ei}{T}}\right),
\end{equation}
where $\nu$ is the oscillation factor  of the CO molecule on the surface, which is typically  $\nu$=10$^{12}$ s$^{-1}$, and T the temperature of the substrate.

\subsection{Theoretical results}

\subsubsection{Deposition}
We have performed simulations for the two different deposition rates used experimentally (1.5 ML/min for CO deposited from 80~K  with decreasing temperatures of 0.5~K/min down to 8~K, and 7.8 ML/min for CO deposited at 14~K). 

\begin{figure*}
    \centering
    \includegraphics[width=0.43\textwidth]{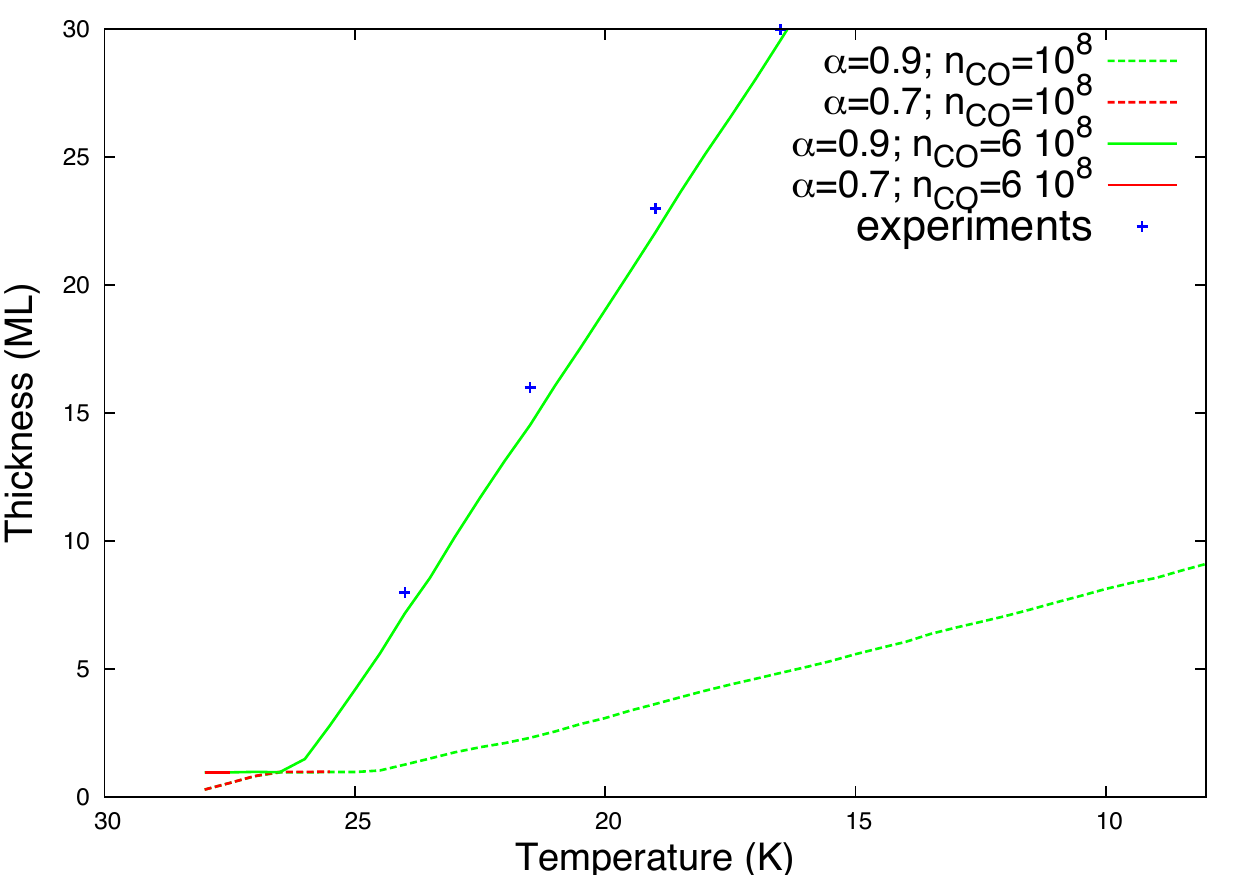}
    \includegraphics[width=0.43\textwidth]{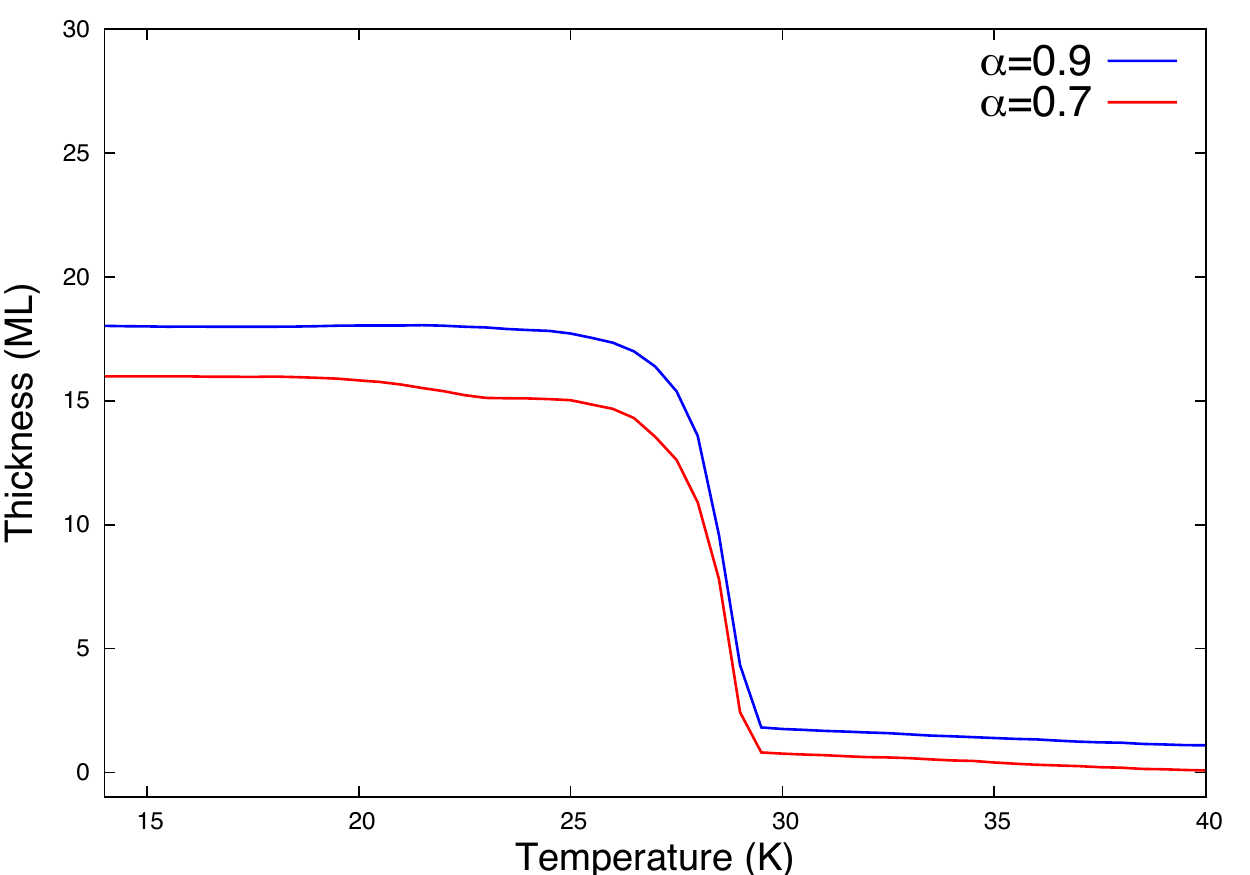}
    \caption{Left panel: Modelled accretion of CO in monolayers as in experiment 1 (crosses). Right panel: modelled thickness evolution as in experiment 2. $\alpha$ shows the different diffusion parameters.}
    \label{stick}
\end{figure*}



\rm{Our first goal is to constrain the diffusion parameter $\alpha$ (ratio between diffusion barrier and binding energy) from the constant accretion measured in experiment 1. In this first experiment, the CO molecules are deposited \rm{on a water ice surface} as the temperature of the substrate is decreased from 80~K to 10~K. In our calculations, the diffusion coefficient $\alpha$ is set to 0.9 and 0.7. Our results presented in the left panel of Fig. \ref{stick} show that in order to reproduce the experimental accretion rate, the diffusion doesn't need to be efficient, since the experimental results are reproduced with $\alpha$=0.9 (low diffusion, see Sect. 3.3). This is because of the very high flow of CO arriving on the surface, needed to reproduce the experimental accretion rate. If the flux was 6 times lower (corresponding to a density of 10$^8$ cm$^{-3}$), then the accretion would be much lower for $\alpha$=0.9. This is because for lower flows, there is a competition between diffusion (which allow to find higher binding sites with more neighbours) and evaporation. Therefore, the first  experiment does not allow to constrain the diffusion parameter.  \\

We then tried to constrain the diffusion parameter  $\alpha$, from the second experiment. For this purpose we computed the evolution of the thickness with increasing temperature, as in the second experiment. Our results, considering a diffusion parameter $\alpha$=0.7 and 0.9, are shown in figure \ref{stick}, right panel. The thickness of the CO ice depends on the diffusion of CO molecules within the ices. If the diffusion barrier is of 90$\%$ of the binding energy ($\alpha$=0.9), then the re-organisation of the CO molecules occurs just before desorption, and this cannot be seen in the change of thickness of the ices. However, for a diffusion barrier of 70$\%$ of the binding energy ($\alpha$=0.7), re-organisation of CO molecules in the ices imply a decrease of the thickness at around $\sim$22~K (which corresponds to $\sim$0.7$\times$ E$_{\rm{bin}}$/30.9 where E$_{\rm{bin}} \sim$900~K is the energy of CO in the bulk), which is not seen in the experiments. In order to observe no thickness decrease between 14 and 27~K, as shown experimentally, the diffusion parameter should be either $\alpha \ge$ 0.9, so that the thickness decrease occurs around ~27~K, or $\alpha$ $<$ 0.5 (T$\sim$0.5$\times$900 /30.9$\sim$ 14.5~K) so that the thickness decrease occurs below 14~K. Therefore, we conclude that CO diffusion within CO ices has to be either lower than 50$\%$ of the binding energy or higher than 90$\%$ of the binding energy. }\rm

\rm{In the experiments, the ices obtained after deposition were heated to higher temperatures and the desorption was measured. To mimic the warming up, we performed simulations with the ices deposited from 80~K to 8~K. To simulate the warming up of the ices, we used conditions similar to the ones met in the first experiment so that  the temperature of the substrate was heated at a rate of 0.5~K/min until 80~K. The TPDs obtained are shown in Fig.~\ref{tpd}. We computed the TPD for two different $\alpha$ of 0.9 and 0.7.  Our simulations can reproduce the multilayer peak located around 30~K as well as the monolayer contribution, extending after  40~K. Also, the small desorption bump occurring at $\sim$20~K in the experiment can be reproduce if the diffusion is inefficient ($\alpha$=0.9). If the diffusion is higher, then the CO molecules do re-organise instead of desorbing. Our model therefore confirms the presence of weakly bound CO molecules in the TPD, that can be explained only if the diffusion of CO in the ice is inefficient. Note that the amount of CO desorbing around $/sim$20~K is small so we used a larger grid of 40$\times$40 to make these computations.}\rm


Our simulations show that even if ices present different structures and different binding energies, the TPD measurements are very similar. However, the subtle differences seen in the TPD, especially at low temperatures, do highlight the different structure of the ice and the presence of weakly bound CO molecules. 
%

\begin{figure} 
    \centering
    \includegraphics[width=0.45\textwidth]{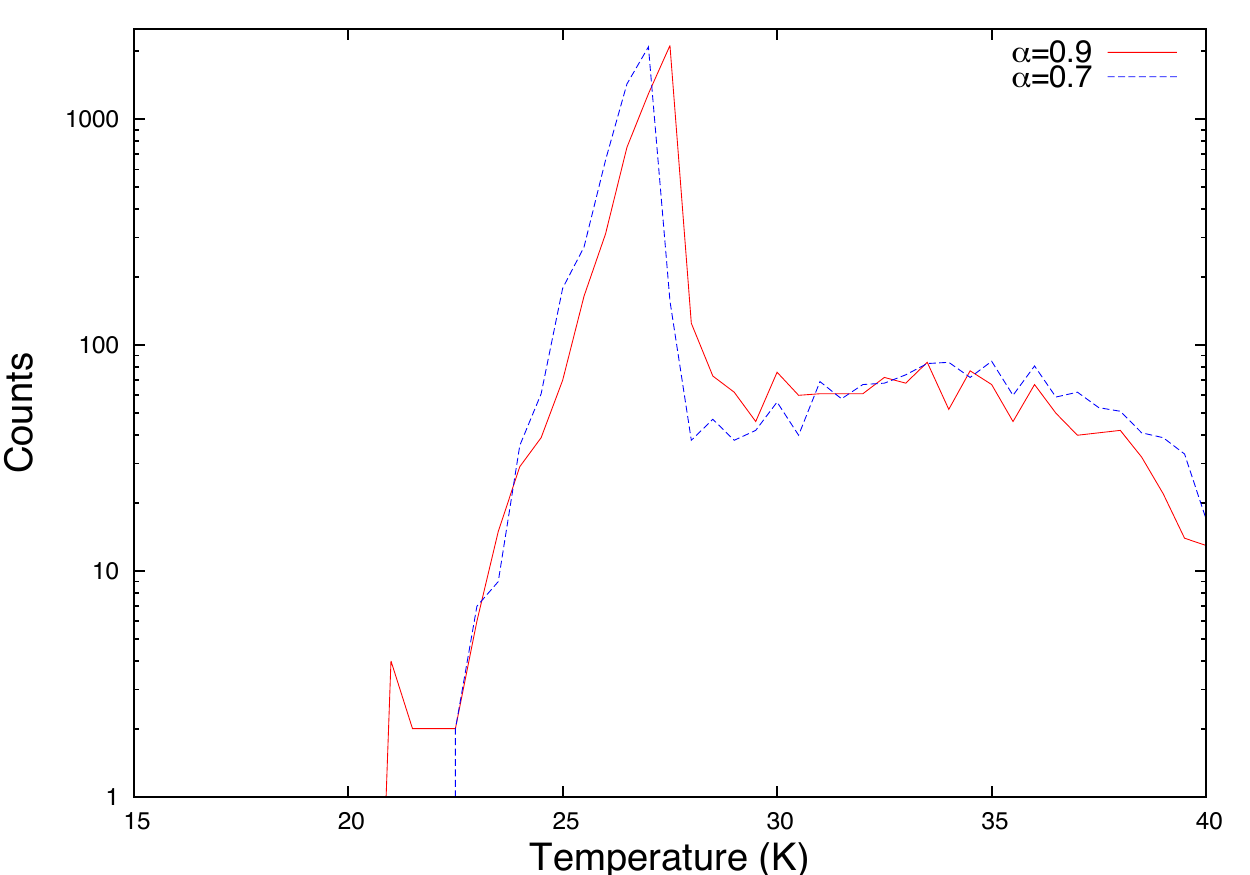}
    \caption{Simulated TPD from Monte Carlo simulations with two different values for $\alpha$. }
    \label{tpd}
\end{figure}

In our simulations, we use a grid of 20 $\times$ 20 sites, and each CO molecule is represented by a box as shown in figures \ref{COads} left panel, for the first experiment and \ref{COads}, right panel, for the second experiment. The different colours show the binding energy of  each of the CO molecules. These binding energies range from 300~K (a CO molecule with one direct neighbour and neighbours on the sides) to $\sim$1700~K (a CO molecule on the deeper sites of the water substrate). 
\begin{figure*}
    \centering
    \includegraphics[width=0.45\textwidth]{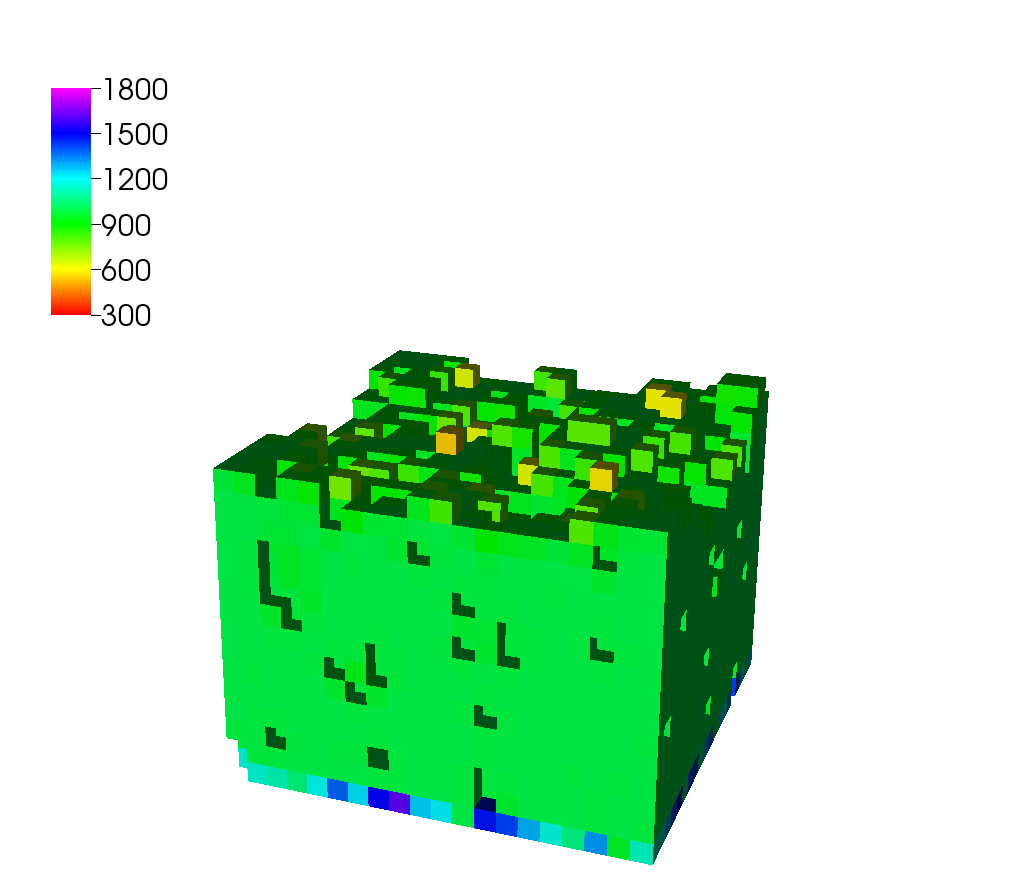}
    \includegraphics[width=0.45\textwidth]{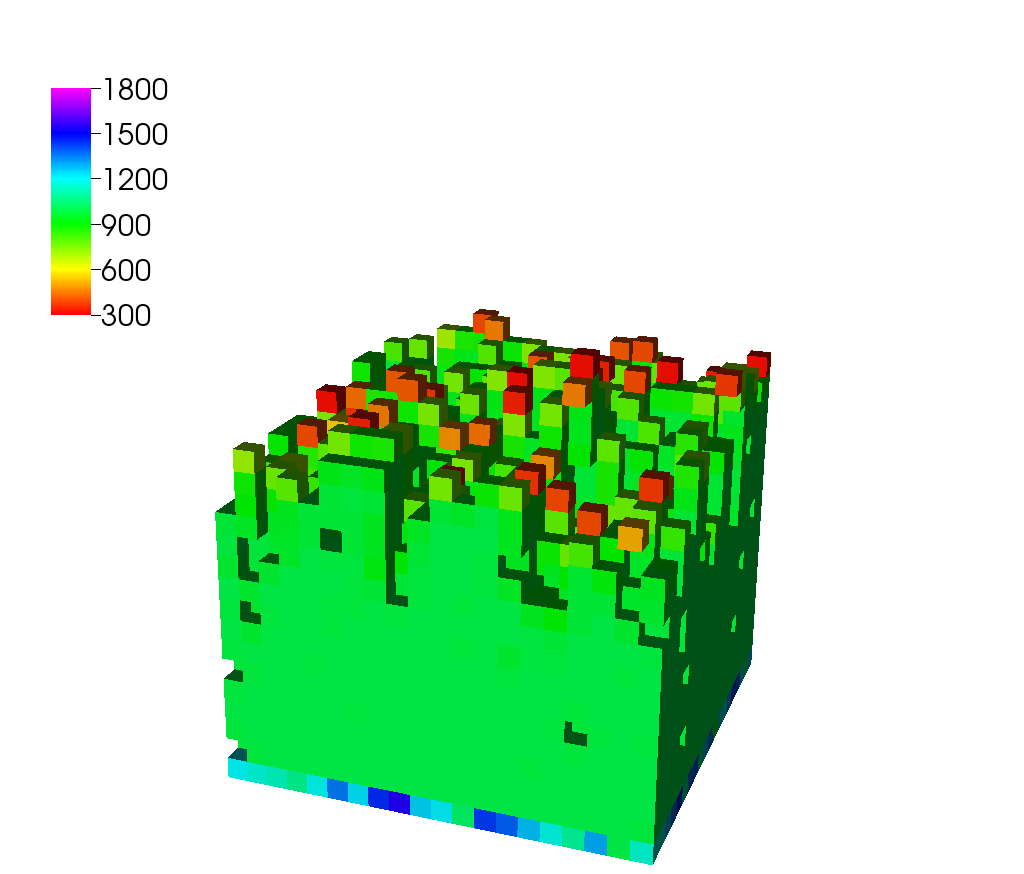}
    \caption{\rm{These figures illustrate the range of binding energies of individual CO molecules after depositing 15 layers. Each square is a CO molecule and the color corresponds to its binding energy in Kelvins. The binding energy of CO molecules on the first level of the grid is higher, because CO molecules are bound to water substrate. Left panel: CO ice on water surface deposited at 14~K. Right panel:  CO ice on water surface deposited from 80~K downwards. This figure shows a hybrid ice, with compact CO ice at the bottom due to deposition at high temperatures and very weakly bound molecules at the top (deposition ends at 8K).}\rm}
    \label{COads}
\end{figure*}

To mimic the first experiment with our simulations, CO molecules are admitted on the surface, and we follow the build up of the CO ice, which takes place layer by layer as the temperature goes down. 
\rm{The first accreted monolayer, formed at temperatures higher than 26.5~K, has no empty sites, since CO molecules can diffuse on the surface to find the binding sites with binding energies high enough so that they can stay on the surface. That is, CO molecules first populate the strongest binding sites.  As the temperature decreases further, the weakest binding sites on the water surface can also be filled.  This makes the first CO layer (directly bound to the water surface) completely filled.}\rm\ 
The drop in gas phase CO at 26.5~K implies that the temperature is low enough to allow the adsorption of CO \rm{molecules} on the first layer of CO. 
For these temperatures, the multilayer regime is reached and CO from the gas phase disappears while CO \rm{ice} can be seen on the surface, as shown in Fig.~\ref{acc}. 
\rm{The next layers of CO are built on a similar manner on top of previously deposited CO (one layer without holes) at 26.5 K. These layers are also very well organised without any holes. As a result, the first layers of ice (deposited at higher temperatures) show almost no holes compared to the ice grown at low  temperatures (see below). As the temperature decreases further, a temperature range is reached at which the molecules do not diffuse anymore. These next layers show more holes and weaker binding CO molecules. The top layer (deposited at 8~K), present many holes and weakly bound molecules. This CO ice is a hybrid ice made of a compact CO ice where no holes can be seen at the bottom layers, and weakly bound molecules and many holes becoming present as the number of layers increase.}\rm\ 

The ice \rm{structure} after deposition in the second experiment is shown in Fig.~\ref{COads}, right panel. In the second experiment, CO molecules are deposited at 14~K on different types of surfaces. In our simulations, we concentrate on the adsorption of CO on water ice.  The resulting CO ice shows many empty spaces, and weakly bound CO molecules. This is due to the fact that the deposition temperature is low enough for the CO molecules to be weakly bound,  \rm{and} not to evaporate. The CO molecules arriving on the surface with low binding energies ($\sim$~200~K)  can go to a site where the binding energy is high enough to settle in that site. 

Therefore, experiments and simulations show that ices deposited at 80~K downwards, and at fixed temperature of 14~K have different structures. We conclude that the deposition temperature sets the binding energy of the CO molecules in the ice.}\rm

\section{Astrophysical applications}

In starless cores, an important CO depletion has been measured (\citealt{bergin2002}). Evidence of depletions by factor of 4-15 (\rm{75-94$\%$ of CO missing from the gas phase}) in many of these cores has been found (\citealt{bacman2002}). However,  by observing CO isotopologues, a \rm{higher} CO depletion of \rm{up to} ~100 and ~1000 in the center of the 
B68 and L1544 \rm{dense clouds}, respectively, \rm{has been} measured ($\ge$99\% of CO missing from the gas phase; \citealt{caselli1999,bergin2002}). 

\begin{figure*}
    \centering
        \includegraphics[width=0.45\textwidth]{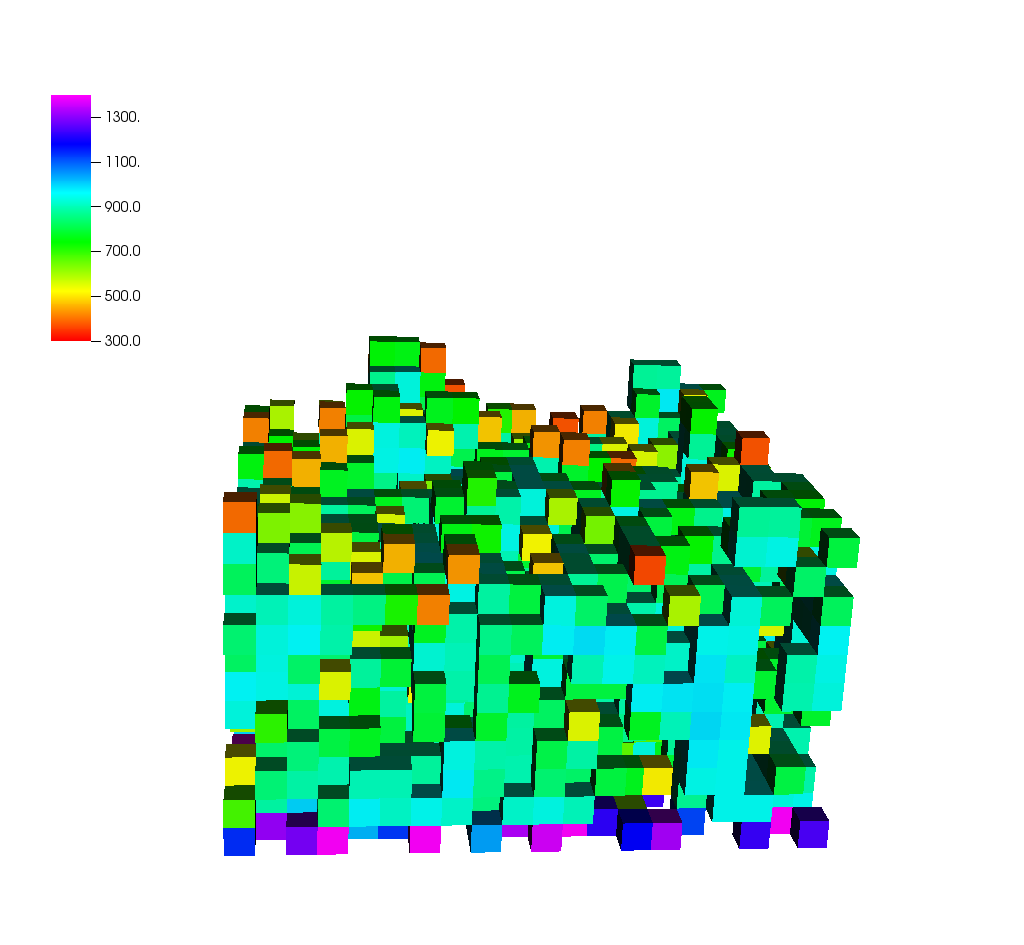}    
    \caption{Binding energy of CO molecules in the CO ice for deposition at T=6 K and at densities of n$_{\rm{H}}$=10$^6$ cm$^{-3}$. The diffusion parameter, $\alpha$ is set to 0.9.}
    \label{binding}
\end{figure*}

The most common explanation is that CO is frozen on dust grains at high densities and low temperatures. However, in order to match the observed CO spectra, \cite{keto2010} found that the desorption rate due to cosmic ray strikes, (\cite{Hasegawa1993}), should be increased by a factor of 30. 
At this rate, desorption and depletion have equal time-scales at a density of about 10$^4$ cm$^{-3}$. 
There are other \rm{non-thermal} processes in addition to direct cosmic ray strikes that cause desorption and could increase the gas-phase abundance of CO. These processes are photodesorption with UV photons (\citealt{oberg2007, oberg2009a, munozcaro2010,fayolle2011, munozcaro2016}); formation of \hm (\citealt{takahashi2000}), non canonical explosions (\citealt{rawlings2013}); direct cosmic ray sputtering (\citealt{dartois2015}) or cosmic ray induced explosive chemical desorption (\citealt{shen2004}); chemical desorption 
(\citealt{dulieu2013}) and impulsive spot heating on grains (\citealt{ivlev2015}). 

In order to understand which processes allow to keep \rm{the required} fraction of the depleted CO in the gas phase, we first compute the binding energies of the CO molecules arriving on the surface with our Monte Carlo simulations. We calculate the binding energies of each CO molecule as they arrive on the CO ices in the dense core conditions (at T$_{\rm{dust}}$=6~K and n$_{\rm{H}}$=10$^{6}$). We here therefore consider only CO in multilayer \rm{regime} (not directly bound to the surface, \rm{but to other CO molecules}). 
\rm{We consider a diffusion parameter $\alpha$ of 0.9 as derived in the experiments. The resulting structure of the CO ice and binding energies of CO molecules in the ices, after moving and being relocalised on the surface is shown in Fig.~\ref{binding}. The CO molecules in the ices are weakly bound and many holes are present. The CO ices deposited in conditions present in pre-stellar cores show many holes and a wide range of binding energies. }\rm

\begin{figure*}
    \centering
    \includegraphics[width=0.55\textwidth]{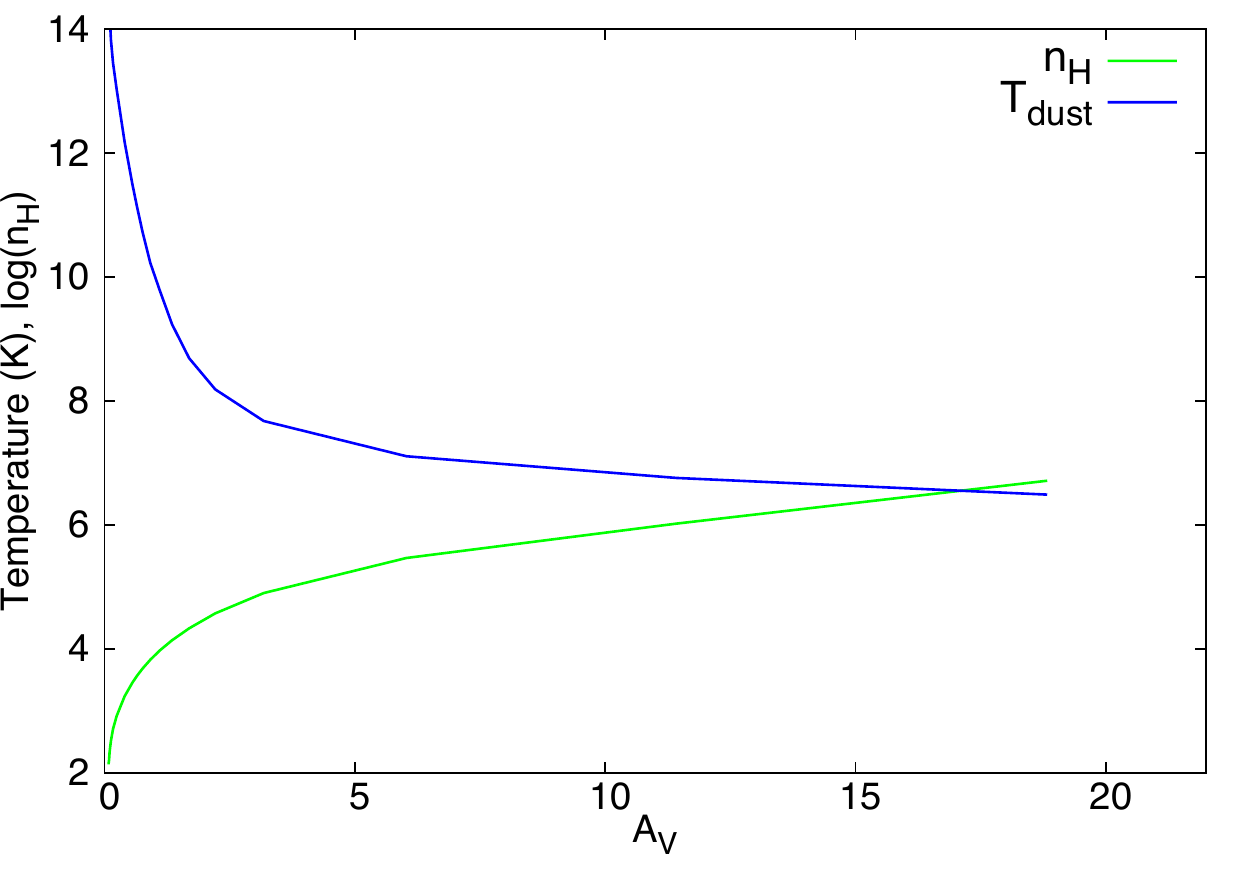}
    \caption{Density temperature profile of pre-stellar cores from \cite{keto2010}.}
    \label{DT}
\end{figure*}

In order to address how \rm{this range of} CO binding energies influence the freeze out of CO molecules from the gas phase, we used a time dependent gas-grain model to follow the abundances of species in the gas, as well as in the ices \rm{in a pre-stellar core}. We used a three-phase chemical model that combines gas-phase chemistry with surface and bulk chemistry. 
The grain surface chemistry model (surface + bulk) takes into account the different binding energies of the species on bare or icy surfaces and includes evaporation, reactions,  photodissociation, and thermal and photodesorption processes, which transform surface species either into other surface species or into gas-phase species, as in \cite{cazaux2016}. 
Note that in this study, we only consider CO freeze out on the dust surface but do not allow surface reactions with CO. 
\rm{In addition,} our model does not take into account the diffusion of species from bulk to surface and from surface to bulk. In this sense, when the coverage has reached one layer, the accreting species become bulk species with higher energies and lower diffusion rates (because the diffusion depends on the binding energy). 
The gas-phase chemical model is adopted from the KIDA database (\citealt{wakelam2012}), while the surface chemistry model is 
described in \cite{cazaux2016}. The input parameters to mimic the temperature and density profile of a pre-stellar core are taken from \cite{keto2010} and shown in Fig.~\ref{DT}.  

\begin{figure*}
    \centering
    \includegraphics[width=0.45\textwidth]{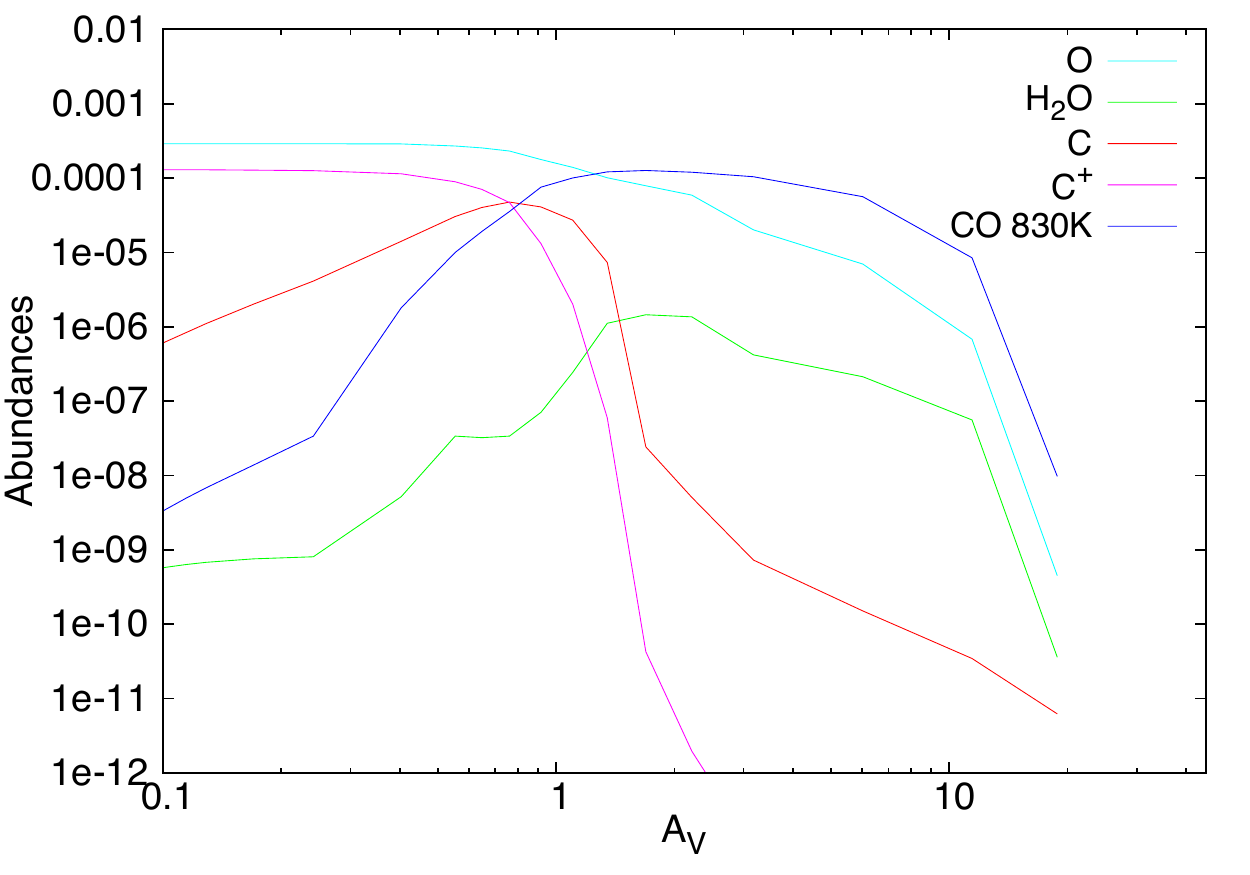}
    \includegraphics[width=0.45\textwidth]{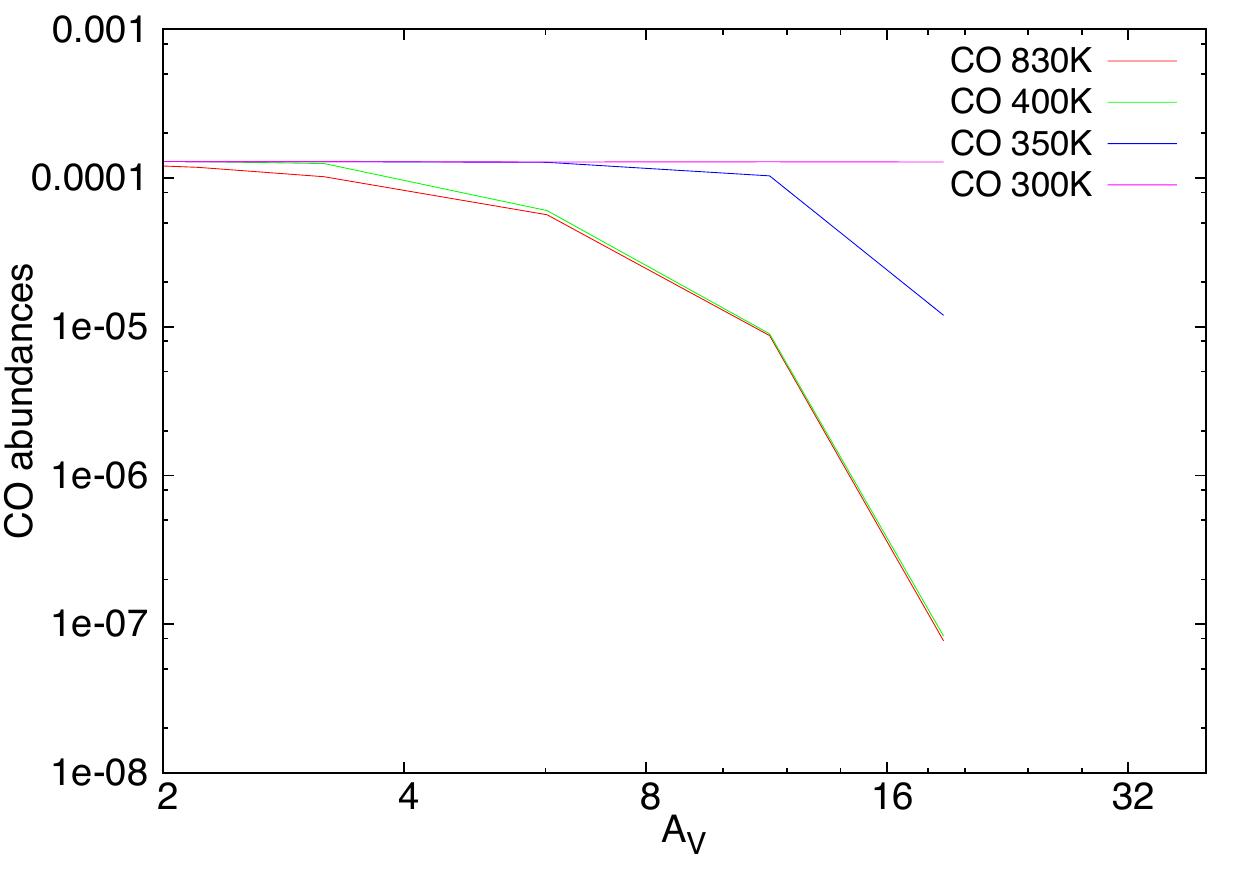}
    \caption{Left: Abundances of C$^+$, C, CO, O and H$_2$O at 10$^4$ years as function of extinction in our pre-stellar core model. Right: CO abundances derived form our model considering binding energies of CO ranging between 300K and 830 K.}
    \label{gas}
\end{figure*}

\begin{figure*}
    \centering
    \includegraphics[width=0.45\textwidth]{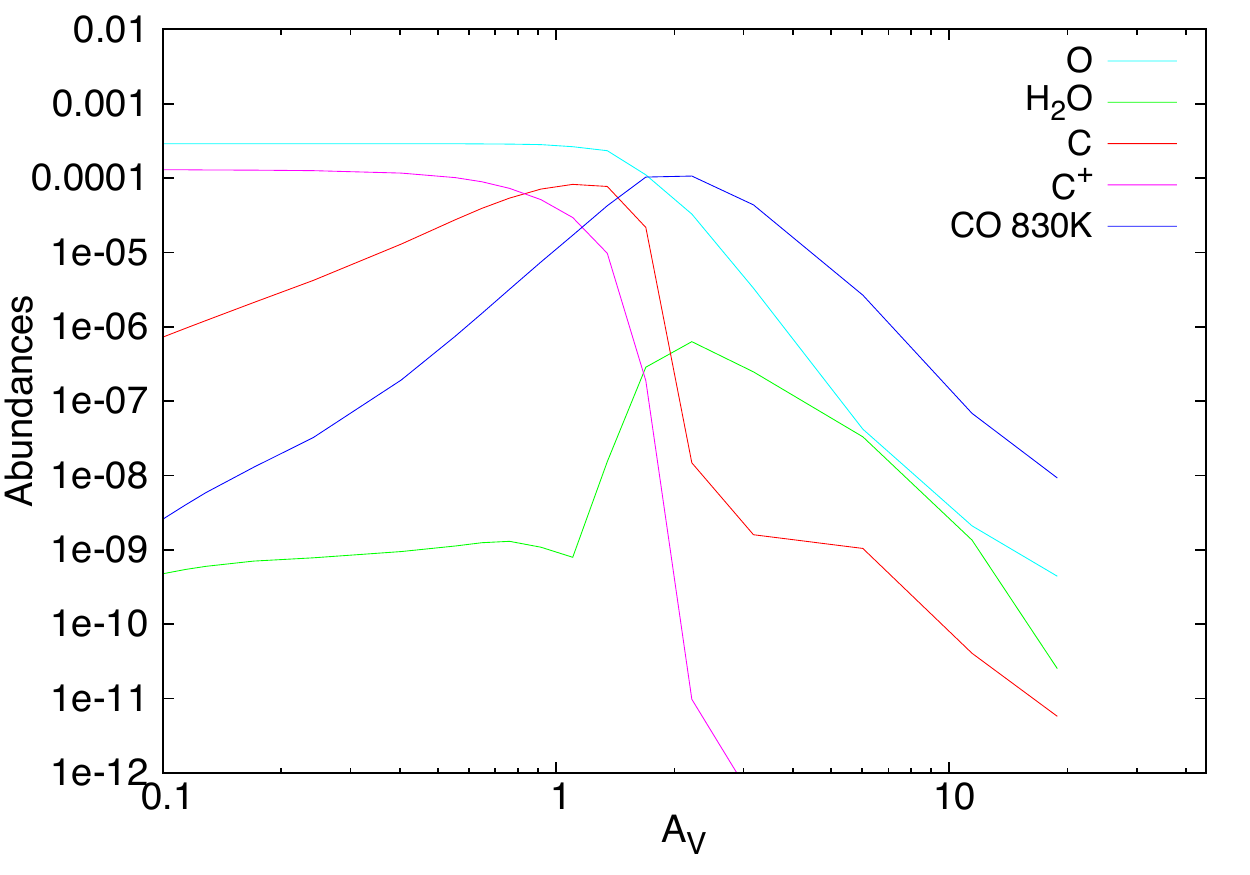}
    \includegraphics[width=0.45\textwidth]{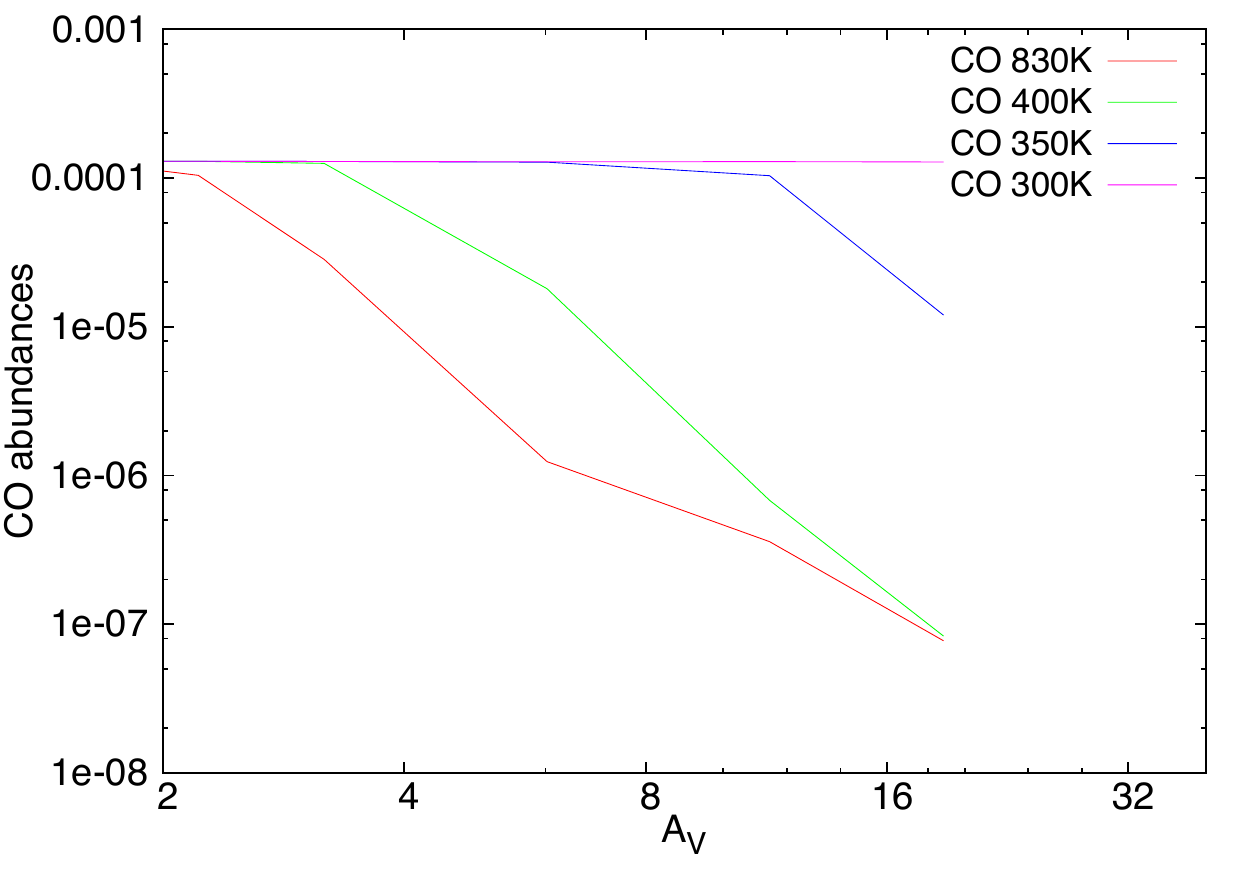}
    \caption{Left: Abundances of C$^+$, C, CO, O and H$_2$O at 10$^5$ years as function of extinction in our pre-stellar core model. Right: CO abundances derived from our model considering binding energies of CO ranging between 300K and 830 K.}
    \label{gas2}
\end{figure*}

The C$^+$, C, CO, O and H$_2$O abundances in the gas phase as a function of extinction are reported in Fig.~\ref{gas} for 10$^4$ 
years, left panel. Our results show that if we consider a binding energy of CO, typical of CO on CO ices $\sim$ 830~K, 
then the CO depletion at extinctions of A$_{\rm{V}}$ $\sim$ 20 can reach 5 orders of magnitude \rm{(0.001\% of CO in the gas phase, 
that is, 99.999\% of CO frozen-out in the ice and missing from the gas phase)}. 
However, as the binding energy of weakly bound CO is considered, such as 400~K (green), 350~K (blue) and 300~K (pink), 
as shown in Fig.~\ref{gas} right panel, the depletion is strongly decreased and CO in the gas phase reaches 0.1$\%$,   10$\%$ and \rm{100}$\%$ 
respectively.

Therefore considering the weakly bound CO molecules can completely change the depletion of CO at high A$_{\rm{V}}$. As the cloud evolves 
and reaches 10$^5$ years, C$^+$, C, CO, O and H$_2$O abundances also evolve as shown in Fig.~\ref{gas2}, left panel. 
For typical binding energies of CO with CO ($\sim$ 830~K), the depletion is more pronounced from extinctions of A$_{\rm{V}}$ $\sim$2. 
When the weakly bound CO molecules are also considered, as shown in Fig.~\ref{gas2} right panel, differences from the depletion 
of CO can be already seen for binding energies of 400~K (green). This is due to the fact that the abundances of CO are set by 
the accretion versus evaporation processes. In this case, the CO bound at low energies will be earlier in equilibrium, which implies that the abundances of CO in the gas phase for bindings of CO of 300 and 350~K do not change between 10$^4$ and 10$^5$ years. However, for CO molecules more strongly bound to the surface, the equilibrium is not reached yet at 400~K and 830~K and depletion is still increasing.

\rm{While we showed that considering low binding energies for CO would change the depletion of CO in dense cores, we could not compute with a rate equation model the depletion of CO due to the distribution of binding energies on the last layer of the ices. To perform such calculations, one should use a more detailed model considering layers to mimic the icy mantle covering dust grains (\citealt{taquet2012}). However, our simulations indicate that in the conditions met in dense cores, the CO depletion could be described by higher CO binding energies at low A$_V$ (between 0 and 4) and lower binding energies after A$_{\rm{V}}$=4. From figures~\ref{gas} and \ref{gas2} right panels, the CO depletion would therefore range between the red and green curves (CO binding between 830 and 400~K) below A$_{\rm{V}}$=4 and would range between the green and blue curves (CO binding between 400 and 350~K) for higher A$_{\rm{V}}$. }\rm

\section{Summary and conclusions}
We showed experimentally that CO ices \rm{can accrete at temperatures below 26.5 K, with an accretion rate that does not depend on the substrate temperature. 
In addition, CO ices deposited under different conditions (decreasing temperatures from $\sim$ 80~K until 8~K versus constant low temperature of 14~K) show TPD spectra with small differences. These differences highlight the two different structure of the CO ices and the presence of weakly bound CO molecules desorbing around $\sim$20K. Using Monte Carlo simulations, we can reproduce such differences and the weakly bound CO molecules at the condition that the mobility is very inefficient. We show that the diffusion barrier should be 90$\%$ of the binding energies to reproduce such experimental results. }\rm

\rm{In this work we 
show that the deposition conditions (flow and temperature) set the binding energy. During the warming up of the CO ices, the CO molecules re-organise in the most 
stable configuration, which result in almost identical TPD spectra. However, while these TPDs appear similar, the subtle differences highlight the differences of the structure of the CO ices. Re-organisation occurs only at temperatures close to desorption. The different structure becomes clear when CO is deposited at 25.5-27~K. At these temperatures closes to the maximum in the desorption during TPD experiments, the CO is deposited mainly as crystalline $\alpha$-CO. The proportion of amorphous-to-crystalline ice decreases as the deposition temperature approaches the temperature of maximum desorption.}


In environments where stars \rm{form}, the temperature can be so low that weakly bound molecules do not re-organise and stay 
weakly bound to dust grains. This has an impact on the gas phase composition of the environment, but also on the chemistry 
occurring on \rm{the surface of dust grains}, which could be more efficient as weakly bound species are more mobile. 
In pre-stellar cores, CO molecules are 
seen to be depleted as the medium becomes denser and cooler. While this is attributed to the freeze out of CO molecules from the 
gas phase, actual models overestimate the freezing of CO on dust. In this work we show that considering weekly and strongly bound 
molecules in the CO ices changes the CO depletion in pre-stellar cores and allow a less severe depletion. 


\acknowledgments
S. C. is supported by the Netherlands Organization for Scientific Research (NWO; VIDI project 639.042.017) and by the European Research Council (ERC; project PALs 320620). This work was supported by the MOST grants MOST 103-2112-M-008-025- MY3 (Y.J.C.) and by the Spanish MINECO under projects AYA2011-29375, AYA2014-60585-P and AYA2015-71975-REDT (R. M.-D. and G. M. M. C.). We acknowledge Dr. Wing-Fai Thi for useful feedback on the manuscript. We would like to thank the anonymous referee for his/her valuable comments which helped to improve the manuscript.

\bibliographystyle{aa}

\end{document}